\documentclass[prc,aps,a4paper,nofootinbib,showkeys,showpacs,twocolumn,floatfix]{revtex4}
\usepackage{graphicx}
\usepackage{hyperref}
\usepackage{amssymb,amsfonts}
\usepackage{amsmath}
\usepackage{epstopdf}
\usepackage{natbib}
\usepackage{color}
\usepackage{subfigure}
\usepackage{nicefrac}
\usepackage{xspace}
\usepackage{mathtools}
\usepackage{wasysym}

\newcommand{\vb}[1]{\boldsymbol{\mathbf {#1}}} 
\newcommand{\mat}[1]{\underline{\vb{#1}}} 
\newcommand{\transpose}{\intercal}
\newcommand{\av}{\vb{a}}
\newcommand{\ev}{\vb{e}}
\newcommand{\kv}{\vb{k}}
\newcommand{\nablav}{\vb{\nabla}}
\newcommand{\Omegav}{\vb{\Omega}}
\newcommand{\phiv}{\vb{\phi}}
\newcommand{\rv}{\vb{r}}
\newcommand{\Sv}{\vb{S}}
\newcommand{\uv}{\vb{u}}
\newcommand{\vv}{\vb{v}}
\newcommand{\xv}{\vb{x}}
\newcommand{\yv}{\vb{y}}
\newcommand{\jvb}{\bar{\vb{j}}}
\newcommand{\vvb}{\bar{\vb{v}}}

\newcommand{\cell}{\mathcal{C}}
\newcommand{\surface}{\Omega}
\newcommand{\ekinn}{\mathcal{E}_{\text{kin},n}}

\newcommand{\In}{2}
\newcommand{\Out}{1}

\newcommand{\nnavg}{\bar{n}_{n}}
\newcommand{\npavg}{\bar{n}_{p}}
\newcommand{\Neff}{N_{\text{eff}}}
\newcommand{\Aeff}{A_{\text{eff}}}

\newcommand{\Nrcluster}{N_{\text{r-cluster}}}
\newcommand{\Arcluster}{A_{\text{r-cluster}}}
\newcommand{\Necluster}{N_{\text{e-cluster}}}
\newcommand{\Aecluster}{A_{\text{e-cluster}}}
\newcommand{\nnboundel}[1]{n^{b}_{n,#1}}
\newcommand{\nnbound}{n^{b}_{n}}
\newcommand{\nnsuper}{n_{n}^{s}}
\newcommand{\nnfree}{n_{n}^{f}}

\newcommand{\nnconfinedin}{n_{n,\In}^{c}}
\newcommand{\nnfreein}{n_{n,\In}^{f}}
\newcommand{\nnmbound}{\mat{n}_n^{b}}
\newcommand{\nnmsuper}{\mat{n}_n^{s}}
\newcommand{\Pcore}{P_{\text{core}}}
\newcommand{\ncore}{n_{\text{core}}}
\newcommand{\Pdrip}{P_{\text{drip}}}
\newcommand{\Icrust}{I_{\text{crust}}}

\newcommand{\UnitM}[1]{\mat{I}_{#1}}

\newcommand{\Eq}[1]{Eq.~(\ref{#1})}
\newcommand{\Eqs}[1]{Eqs.~(\ref{#1})}
\newcommand{\Ref}[1]{Ref.~\cite{#1}}
\newcommand{\Refs}[1]{Refs.~\cite{#1}}
\newcommand{\Fig}[1]{Fig.~\ref{#1}}
\newcommand{\Sec}[1]{Sec.~\ref{#1}}
\newcommand{\laplace}{\mathop{}\!\mathbin\bigtriangleup}

\begin{document}
  \title{Superfluid hydrodynamics in the inner crust of neutron stars}
  \author{No\"el Martin}
  \email{noelmartin@ipno.in2p3.fr}
  \author{Michael Urban}
  \email{urban@ipno.in2p3.fr}

  \affiliation{Institut de Physique Nucl\'eaire, CNRS/IN2P3,
    Universit\'e Paris-Sud, and Universit\'e Paris-Saclay, F-91406
    Orsay Cedex, France} 

  \begin{abstract}
    The inner crust of neutron stars is supposed to be inhomogeneous
    and composed of dense structures (clusters) that are immersed in a
    dilute gas of unbound neutrons. Here we consider spherical
    clusters forming a BCC crystal and cylindrical rods arranged in a
    hexagonal lattice. We study the relative motion of these dense
    structures and the neutron gas using superfluid
    hydrodynamics. Within this approach, which relies on the
    assumption that Cooper pairs are small compared to the crystalline
    structures, we find that the entrainment of neutrons by the
    clusters is very weak since neutrons of the gas can flow through
    the clusters. Consequently, we obtain a low effective mass of the
    clusters and a superfluid density that is even higher than the
    density of unbound neutrons. Consequences for the constraints from
    glitch observations are discussed.
  \end{abstract}

  \maketitle
\section{Introduction}
The inner crust of neutron stars is characterized by the presence of
clusters in a more dilute gas of unbound neutrons. While the clusters,
containing protons and neutrons, form probably a periodic lattice in
order to minimize the Coulomb energy, the neutron gas is
superfluid. The superfluid component of the crust can have potentially
observable consequences for the hydrodynamical and thermodynamical
properties of the crust \cite{PageReddy2012}. It is therefore
important to know the density of effectively free neutrons. This is a
non-trivial problem because even the unbound neutrons might be
``entrained'' by the clusters because of their interactions.

This entrainment effect has already been extensively discussed in the
literature \cite{PrixComer2002}, mostly  in the framework
of a band-structure theory for neutrons developed by Chamel and
co-workers \cite{CarterChamel2005,Chamel2006,Chamel2012}. This theory
predicts that a large fraction of the free neutrons are
entrained. Consequently, the density of effectively free superfluid
neutrons is strongly reduced. However, as discussed in
\Refs{AnderssonGlampedakis2012,Chamel2013PRL}, it is
difficult to conciliate this reduction of the superfluid density with
the observed glitch activity of the Vela pulsar.

The entrainment has also a strong effect on the heat transport
properties of the crust, and consequently on the cooling of
the star, through a modification of the speed of lattice and
superfluid phonons \cite{ChamelPage2013,KobyakovPethick2013}. These
have been discussed in the framework of an effective theory for
low-energy, long-wavelength excitations \cite{CiriglianoReddy2011}. A
long wavelength means in this context a wavelength that is large
compared to the periodicity of the crystalline structures in the
crust. This effective theory has a couple of parameters that have to
be determined from more microscopic approaches. Among these parameters
are the effective masses of the clusters, or, equivalently, the
superfluid density.

However, under the assumption that pairing is sufficiently strong,
superfluid hydrodynamics can also be applied on length scales that are
smaller than the periodicity of the crystalline structures. This idea
was used in
\cite{Sedrakian1996,MagierskiBulgac2004,MagierskiBulgac2004NPA,Magierski2004}
to estimate the effective mass of an isolated cluster immersed in a
neutron gas, and more recently also to describe collective modes in
the so-called ``lasagne'' phases in the deepest layers of the inner
crust \cite{DiGallo2011,UrbanOertel2015}. In the present work, we
apply this superfluid hydrodynamics approach also to the crystalline
and ``spaghetti'' phases.

In \Sec{sec:model}, we briefly summarize the hydrodynamic model and
the underlying assumptions. Then we apply it to the uniform motion of
a crystalline lattice of clusters relative to the neutron gas
(\Sec{sec:uniform-flow}) and discuss how it can be related to the
macroscopic entrainment (\Sec{sec:entrainment}). In
\Sec{sec:solution}, we explain how the hydrodynamic equations are
solved. The properties of the specific geometries, namely the
Body-Centered Cubic (BCC) crystal of spherical clusters and the
hexagonal lattice of cylindrical rods, are discussed in
\Sec{sec:geometries}. Numerical results for microscopic and
macroscopic quantities and consequences for the interpretation of
glitches are presented in \Sec{sec:results}. We conclude with a
discussion in \Sec{sec:discussion}.

Except in \Sec{sec:glitches} and in the appendix, we use units with
$\hbar = c = 1$, where $\hbar$ is the reduced Planck constant and $c$
the speed of light.
\section{Formalism}
\label{sec:formalism}
\subsection{Hydrodynamic model for the inner crust}
\label{sec:model}
Let us briefly recall the simple hydrodynamic model of
\Refs{MagierskiBulgac2004,MagierskiBulgac2004NPA,Magierski2004,UrbanOertel2015}. We
assume that the clusters have constant neutron and proton densities
$n_{n,\In}$ and $n_{p,\In}$ and a sharp suface separating them from
the neutron gas, whose density $n_{n,\Out}$ is also constant. The
densities have to satisfy the conditions of phase equilibrium (equal
chemical potentials and pressure in both phases), which is actually a
very good approximation \cite{MartinUrban2015}.

Furthermore, it is assumed that the neutrons are
superfluid. Therefore, low-energy excitations correspond to coherent
flow of Cooper pairs. If the superfluid order parameter (gap) is
written as $\Delta = |\Delta|e^{i\varphi}$, the velocity field of the
neutron pairs is related to the phase $\varphi$ by $\vv_n =
\nablav\varphi/(2m)$, where $m$ denotes the neutron mass.\footnote{In
  contrast to \Refs{DiGallo2011,UrbanOertel2015} we neglect here the
  ``microscopic'' entrainment of neutrons by protons in the liquid
  phase \cite{BorumandJoynt1996}, which originates from the velicity
  dependence of the effective neutron-proton interaction. It should be
  included in future studies.} In the limit of zero temperature, and
if one excludes pair breaking, this leads to the equations of
superfluid hydrodynamics as discussed in
\Refs{UrbanSchuck2006,ToniniWerner2006} in the context of ultracold
atoms. Let us also mention that, again in the context of ultracold
atoms, a calculation in quasiparticle random-phase approximation
(QRPA) \cite{GrassoKhan2005} showed that the collective modes can be
described by hydrodynamics if $|\Delta|$ becomes much larger than the
spacing of the discrete single-particle levels in the trap potential.

In uniform neutron matter, the QRPA shows that the hydrodynamic
behavior of the oscillations of the phase $\varphi$ (Goldstone or
Bogoliubov-Anderson mode) is well fulfilled as long as the excitation
energy stays well below the two-quasiparticle (pair breaking)
threshold \cite{MartinUrban2014}. Furthermore, QRPA calculations
of collective modes of a cluster in a spherical Wigner-Seitz (WS) cell
predicted the appearance of ``supergiant'' resonances that could be
interpreted as hydrodynamic Bogoliubov-Anderson modes in the volume of
the cell \cite{KhanSandulescu2005}.

\begin{figure}
  \includegraphics[width=8cm]{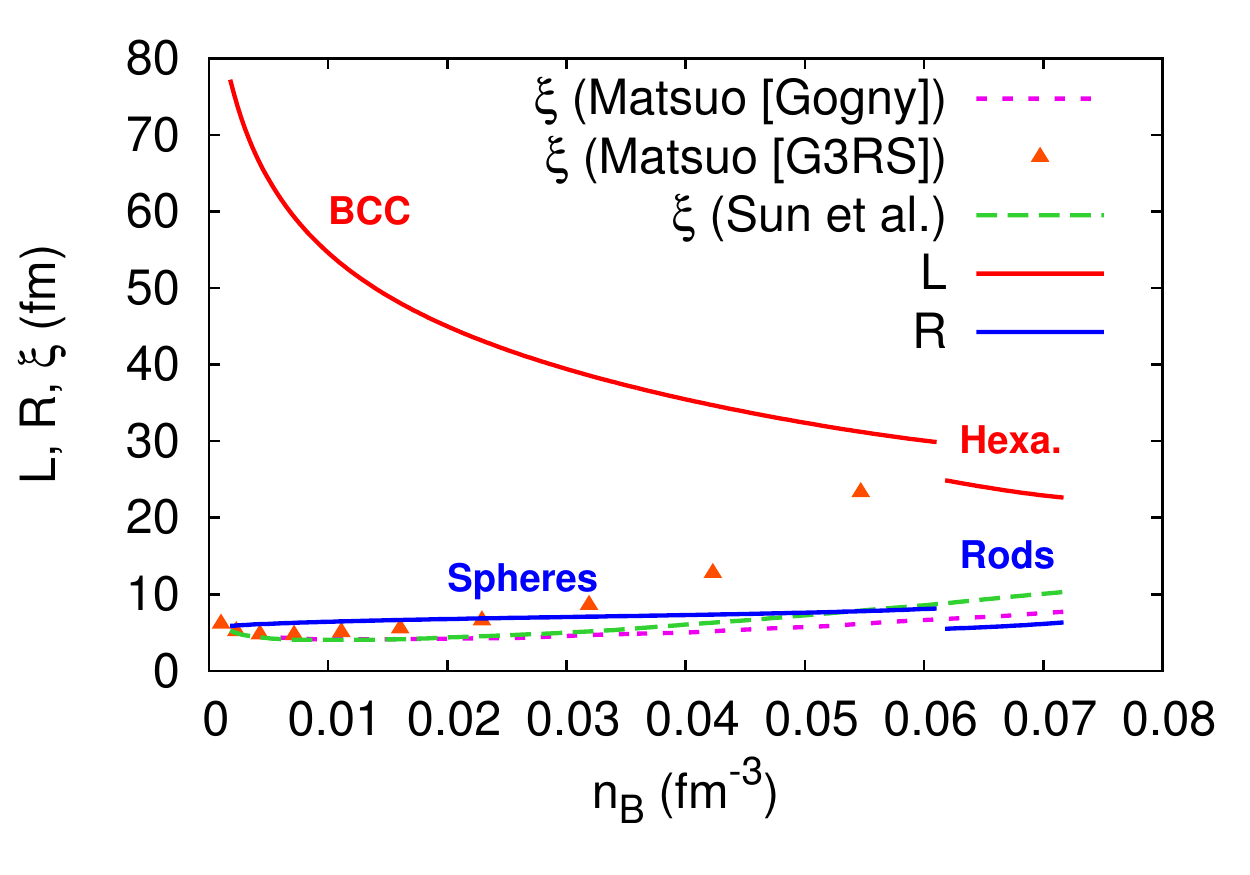}
  \caption{(Color online) Root mean square radius $\xi$ of the Cooper
    pair in the neutron gas (dashed lines) compared with the cell size
    $L$ of the crystalline lattice (red solid line) and the cluster
    radius $R$ (blue solid line) as functions of the total baryon
    density $n_B$ in the inner crust. The neutron gas density
    $n_{n,\Out}$ and the shown results for $L$ and $R$ were obtained
    from calculations detailed in \Ref{MartinUrban2015}. The results
    for $\xi$ as functions of $n_{n,\Out}$ were obtained respectively
    by Matsuo \cite{Matsuo2006} and Sun et al. \cite{SunToki2010}
    using the Gogny force (purple short dashes), the G3RS force
    (orange triangles), and the Bonn potential (green long dashes) as
    pairing interactions.
\label{fig:L-R-xi}}
\end{figure}

In a non-uniform system, hydrodynamics is valid if the coherence
length $\xi$ of the Cooper pairs is small compared to the size of the
inhomogeneities. In \Fig{fig:L-R-xi},
we show different theoretical predictions of the Cooper pair size
$\xi$ in the neutron gas. While the results obtained with the Gogny
force \cite{Matsuo2006} are in good agreement with those obtained
with the Bonn potential \cite{SunToki2010}, the coherence length
obtained with the G3RS force \cite{Matsuo2006} is considerably
larger, especially at higher densities $n_{n,\Out}$. Actually, the
uncertainty in $\xi$ is directly related to the fact that the
density dependence of the gap $\Delta$ in neutron matter is not very
well known. From now on we will assume that pairing is strong, as
with the Gogny or Bonn interactions. Note that the coherence length
is also relevant for the spatial structure of vortices
\cite{ElgaroyDeBlasio2001,YuBulgac2003}.

In \Fig{fig:L-R-xi}, we also display the size $L$ of the unit cell of
the crystalline lattice and the radius $R$ of the clusters obtained in
\Ref{MartinUrban2015}. Although the exact numbers for $L$ and $R$
depend on the model, their order of magnitude follows from the balance
between surface and Coulomb energy and is therefore much better
constrained than the coherence length. We see that, at low densities,
the Cooper pair size $\xi$ in the neutron gas is indeed much smaller
than the size $L$ of the unit cell of the crystalline lattice. At
higher densities, where one expects the pasta phases, the comparison
would be somewhat less favorable. However, the main problem is the
small size of the clusters, whose radii $R$ are as small as $\xi$ even
in the case of strong pairing.

The condition $\xi \ll R$ for the validity of hydrodynamics was
already mentioned in Migdal's seminal work \cite{Migdal1959} in which
he explained the nuclear moments of inertia in the framework of the
theory of superfluidity. Since the coherence length $\xi$ and the
nuclear radius $R$ are of the same order of magnitude, rotating nuclei
exhibit a combination of irrotational and rotational
flow. Nevertheless, the nuclear moments of inertia are slightly closer
to the irrotational (hydrodynamic) than to the rigid-body limit (see
Fig. 8.2 in \Ref{Rowe}).

In analogy to this observation, we expect that probably our superfluid
hydrodynamic model for the inner crust should give the right picture,
although it might probably overestimate the superfluid flow inside
(and close to) the clusters. We refer to \Sec{sec:discussion} for a
further discussion of this problem and possible solutions.
\subsection{Uniform flow of clusters through the gas}
\label{sec:uniform-flow}
In the present paper, we concentrate on macroscopic (long wavelength)
motion. In this case, the relative velocity between clusters and
neutron gas varies only on length scales that are much larger than the
periodicity of the lattice.

We limit ourselves to a stationary motion, i.e., we assume that
the velocities and densities are time-independent in the rest frame of
the clusters. Then, in the limit of small velocities, the size
and shape of the clusters themselves as well as the densities in the
clusters and in the gas do not change as compared to the static
case. We define the cluster surface as the surface of the
sphere (3D) or rod (2D) containing the protons. Hence, the velocity of
the clusters is equal to the velocity of the protons, $\uv_p$. The
neutrons, however, can pass through the cluster surface, and their
velocity field $\vv_n(\rv)$ is not uniform, since the neutrons of the
gas somehow have to flow around or through the clusters. To determine
this motion is the main goal of the present work.

As mentioned in \Sec{sec:model}, the superfluidity of the neutron gas
allows us to introduce a velocity potential $\phi =
\varphi/(2m)$. Since the densities remain constant, we have
$\nablav\cdot\vv_n = 0$, i.e.,
\begin{equation}
  \laplace \phi = 0 \, .
  \label{eq:laplace-general}
\end{equation}
This equation is true in both phases, but it has to be complemented
with suitable boundary conditions at the phase boundaries.

In \Refs{Sedrakian1996,DiGallo2011}, the phase boundary was treated as
impermeable. However, this is not realistic, since neutrons inside and
outside the cluster are indistinguishable and nothing prevents them
from moving from the gas into the cluster or vice versa. The
permeability of the phase boundary was included in the boundary
conditions introduced by Magierski and Bulgac
\cite{MagierskiBulgac2004,MagierskiBulgac2004NPA,Magierski2004}. Analogous
boundary conditions were given in \Ref{Lazarides2008} for a phase
boundary in the context of ultracold atoms. They were also used in
\Ref{UrbanOertel2015} to describe collective modes in the ``pasta''
phases of the neutron-star crust.

First, the phase of the order parameter is continuous across the phase
boundary, i.e.,
\begin{equation}
\phi_\Out = \phi_\In\,, \label{eq:boundary-condition1}
\end{equation}
where $\Out$ and $\In$ refer to the limits of $\rv$ approaching the
interface from outside or inside the cluster, respectively. This
boundary condition implies that the neutron velocity tangential to the
interface is continuous, too.

Second, the neutron current crossing the interface conserves the
particle number. Since the interface itself moves with velocity
$\uv_p$, this condition reads
\begin{equation}
    n_{n,\Out}(\nablav \phi_{\Out} - \uv_p) \cdot \Sv =
    n_{n,\In} (\nablav \phi_{\In} - \uv_p) \cdot \Sv \,,
  \label{eq:boundary-condition2}
\end{equation}
where $\Sv$ is the normal vector to the surface, pointing outwards.
Note that in the limiting case of a vanishing gas density
($n_{n,\Out}=0$), this equation implies that $\nablav\phi_\In =
\uv_p$, i.e., in this case the neutrons inside the cluster move
together with the protons as is intuitively clear.

So far, the boundary conditions are the same as in
\Ref{MagierskiBulgac2004,MagierskiBulgac2004NPA,Magierski2004}, where
the motion of a spherical nucleus in an infinite neutron gas was
studied. In this case,
\Eqs{eq:laplace-general}--(\ref{eq:boundary-condition2}) can be solved
analytically (see Sec.~\ref{sec:analytic}). However, except in the
case of plates (1D), this is no longer true if one considers a
periodic lattice of clusters.

To treat the periodicity, we introduce a primitive cell $\cell$
spanned by the $D$ primitive vectors $\av_i$ ($i=1,\dots,D$) of the
Bravais lattice, where $D=3$ in the case of a crystal, $D=2$ in the
case of rods (spaghetti phase), and $D=1$ in the case of plates
(lasagne phase). Depending on the lattice structure, the primitive
cell contains one or two clusters (see \Sec{sec:geometries}). While
the velocity field $\vv_n(\rv)$ is periodic,
\begin{equation}
\vv_n(\rv+\av_i) = \vv_n(\rv)\,,
\end{equation}
the velocity potential itself can in general be the sum of a periodic
and a linear function. The linear function can be written as
$\uv_n\cdot\rv$, where $\uv_n$ is the spatially averaged neutron
velocity, which coincides with the velocity of the \emph{superfluid}
neutrons \cite{PethickChamel2010,KobyakovPethick2013} or
\emph{conduction} neutrons \cite{ChamelPage2013}. Note that $\uv_n$ is
different from the average neutron velocity $\bar{\vv}_n$, which is
defined via the spatially averaged neutron current (see
below). Without loss of generality, let us choose the frame of
reference such that $\uv_n = 0$. In this frame, also the velocity
potential is periodic,
\begin{equation}
\phi(\rv+\av_i) = \phi(\rv)\,.
\label{eq:boundary-condition3}
\end{equation}

From the function $\phi(\rv)$ in the primitive cell one can derive the
macroscopic (coarse grained) neutron current $\jvb_n$ by averaging over
the volume of the cell, $V_\cell$:
  \begin{align}
    \jvb_n = \frac{1}{V_\cell} \int_{\cell} dV \,n_n(\rv)
    \nablav{\phi(\rv)} \, .
    \label{eq:jmean-expanded}
  \end{align}
After integration by parts, \Eq{eq:jmean-expanded} reduces to
  \begin{equation}
    \jvb_n = \frac{1}{V_\cell} (n_{n,\In} - n_{n,\Out})
      \oint_{\surface} d\Sv \, \phi(r) \, ,
    \label{eq:jmean}
  \end{equation}
where $\surface$ is the surface of the cluster(s) in the cell. The
integral over the cell boundary vanishes because of the periodicity of
$\phi$.

Similarly, one can calculate the average kinetic energy density
\begin{equation}
  \ekinn = \frac{m}{2V_\cell} \int_{\cell} dV \, n_n(\rv)
     [\nablav \phi(r)]^2 \, .
  \label{eq:tau-expanded}
\end{equation}
Using the Gauss theorem and \Eq{eq:boundary-condition2}, this
expression can be simplified to \cite{MagierskiBulgac2004}
\begin{equation}
  \ekinn = \frac{m}{2V} (n_{\In} - n_{\Out})
    \oint_{\surface} d\Sv \cdot \uv_p \phi(\rv)
     = \frac{m}{2} \uv_p \cdot \jvb_n \,.
  \label{eq:tau}
\end{equation}
\subsection{Entrainment}
\label{sec:entrainment}
In \Eq{eq:boundary-condition3} we assumed that $\uv_n = 0$. The
solution for $\phi$ in the general case $\uv_n\neq 0$ is related to
the periodic solution in the special case $\uv_n = 0$ by
\begin{equation}
  \phi(\rv;\uv_p,\uv_n) = \rv \cdot \uv_n + \phi(\rv;\uv_p - \uv_n,0)\,.
    \label{eq:Phi_def}
\end{equation}
The average velocity of neutrons $\vvb_n$ is defined via the average
current $\jvb_n$ as
\begin{equation}
\vvb_n = \frac{\jvb_n}{\nnavg}\,,
\end{equation}
where
\begin{equation}
\nnavg = \frac{V_\Out}{V_\cell}
n_{n,\Out}+\frac{V_\In}{V_\cell} n_{n,\In}
\end{equation}
denotes the average neutron density with $V_{\Out,\In}$ the volume
outside and inside the cluster(s), respectively. The neutron current
is now written as
\begin{multline}
  \jvb_n = \frac{1}{V_\cell} \int_{\cell} dV \, n_n(r)
    \nablav \phi(\rv;\uv_p,\uv_n)\\
     = \nnavg \uv_n + \frac{1}{V_\cell} \int_{\cell} dV \, n_n(r) \,
       \nablav \phi(\rv;\uv_p - \uv_n,0) \, .
  \label{eq:jm_reduced}
\end{multline}
Since the last term in \Eq{eq:jm_reduced} is linear in $\uv_p-\uv_n$,
we can write the current in the form
\begin{equation}
  \jvb_n = \nnavg \uv_n + \nnmbound (\uv_p - \uv_n) \, ,
  \label{eq:current_matrix}
\end{equation}
with a $3\times 3$ matrix $\nnmbound$. Factorizing
\Eq{eq:current_matrix} with respect to $\uv_n$, one sees that
$\nnmbound$ can be interpreted as the density of bound neutrons, which
are entrained by the clusters with velocity $\uv_p$, while the
superfluid neutrons moving with velocity $\uv_n$ have an average
density $\nnmsuper = \nnavg \UnitM{3} - \nnmbound$. Concerning
bound and superfluid neutrons, we follow here the nomenclature of
\Ref{PethickChamel2010}. Hence, the final expression for the neutron
current reads:
\begin{equation}
  \jvb_n = \nnmbound \uv_p + \nnmsuper \uv_n \, .
\end{equation}

The fact that $\nnmbound$ and $\nnmsuper$ are matrices shows that the
proportion of bound neutrons depends in general on the direction of
the relative motion between neutrons and protons. This is intuitively
clear, e.g., in the case of the 2D rod phase, where neutrons and
protons can move independently of each other in the direction parallel
to the rods, while this is not the case in the directions
perpendicular to the rods. As will be shown in \Sec{sec:bcc},
$\nnmbound$ and $\nnmsuper$ are proportional to the unit matrix if the
lattice has a cubic symmetry.

It is straight-forward to generalize also \Eq{eq:tau} for the neutron
kinetic energy to the general case $\uv_n\neq 0$. First, note that in
the case $\uv_n = 0$, the current simplifies to $\jvb_n = \nnmbound
\uv_p$, and consequently \Eq{eq:tau} becomes $\ekinn = (m/2)
\uv_p^\transpose \nnmbound \uv_p$. Starting from \Eq{eq:Phi_def} and
repeating the same steps for the general case $\uv_n \neq 0$, one
obtains:
\begin{equation}
\ekinn = \frac{m}{2} \left( \uv_n^\transpose \nnmsuper \uv_n +
  \uv_p^\transpose \nnmbound \uv_p \right) \, ,
\label{ekindiagonal}
\end{equation}
which agrees with the expression of Chamel and Carter
\cite{ChamelCarter2006} if one identifies $\nnbound$ with the neutron
\emph{normal} density in their nomenclature.

In summary, the macroscopic entrainment coefficients of the crust are
determined by the matrices $\nnmbound$ and $\nnmsuper$ which we can
obtain by solving numerically
\Eqs{eq:laplace-general}--(\ref{eq:boundary-condition3}) for the
function $\phi(\rv;\uv_p,0)$.

\section{Solution of the hydrodynamic equations}
\label{sec:solution}
\subsection{Analytic solution in simple cases}
\label{sec:analytic}
In the case of a single cluster (spherical or cylindrical) moving with
velocity $\uv_p$ through in an infinite and uniform neutron gas, and
in the case of a 1D lattice of parallel plates, analytical solutions
for the velocity potential can be found.

The case of a spherical cluster of radius $R$ was studied in
\Refs{MagierskiBulgac2004,MagierskiBulgac2004NPA,Magierski2004}. If we
place the origin of the coordinate system in the center of the cluster
and suppose that the neutron gas is at rest at infinity ($\phi\to 0$
for $r\to\infty$), the solution for the velocity potential is
\begin{equation}
  \phi(\rv) =
  \begin{dcases}
    \frac{1-\gamma}{1+2\gamma}\, \rv \cdot \uv_p & \text{for~} r < R \,,
    \\
    \frac{R^3}{r^3} \, \frac{1-\gamma}{1+2\gamma} \, \rv \cdot \uv_p &
    \text{for~} r \geq R \,,
  \end{dcases}
  \label{phi-spherical}
\end{equation}
where $\gamma = n_{n,\Out} / n_{n,\In}$ is the ratio between the
neutron densities in the gas and in the cluster. From this solution,
one can compute the total momentum carried by neutrons in the cluster
and in the gas. Identifying this momentum with $\Neff m\uv_p$, one can
define the number $\Neff$ of neutrons effectively entrained by the
protons of the cluster,
\begin{equation}
  \Neff = \Nrcluster \frac{(1-\gamma)^2}{1 + 2\gamma} \,,
  \label{NeffMagierski}
\end{equation}
with
\begin{equation}
  \Nrcluster = \frac{4 \pi}{3} R^3 \, n_{n,\In} \label{Nrcluster}
\end{equation}
the number of neutrons that are located inside the cluster (in
coordinate space, denoted r-cluster following \Ref{Papakons2013}). It
is interesting to note that $\Neff < \Nrcluster$, i.e., the main
effect is not that the cluster entrains neutrons of the gas with it,
but rather that the flow of gas neutrons through the cluster surface
reduces the speed of the neutrons inside the cluster.

The case of a cylindrical rod moving through an infinite and uniform
neutron gas can be treated analogously. Here, the velocity potential
is given by
\begin{equation}
  \phi(\rv) =
  \begin{dcases}
    \frac{1-\gamma}{1+\gamma}\, \rv_\perp \cdot \uv_p & \text{for~}
      r_\perp < R \,, \\
    \frac{R^2}{r_\perp^2}\,\frac{1-\gamma}{1+\gamma}\,\rv_\perp\cdot\uv_p&
    \text{for~} r_\perp \geq R \,,
  \end{dcases}
  \label{phi-spaghetti}
\end{equation}
where $\rv_\perp$ is the projection of $\rv$ on the plane
perpendicular to the symmetry axis of the rod. Since the rod is
assumed to be infinite, one can only define $\Neff$ and $\Nrcluster$
as numbers per unit length, e.g., $\Nrcluster = \pi R^2 n_{n,\In}$. If
the proton velocity $\uv_p$ is parallel to the rod, the surface of the
rod does not move and there is obviously no entrainment. However, for
$\uv_p$ perpendicular to the rod, the expression of effectively bound
(entrained) neutrons reads as
\begin{equation}
  \Neff = \Nrcluster \frac{(1-\gamma)^2}{1+\gamma} \,.
  \label{Neff-spaghetti}
\end{equation}
One sees that the number of entrained neutrons is again lower than the
number of neutrons geometrically located inside the rod.

Another case in which an analytic solution can be found is the phase
of plates (1D). Let us consider alternating layers of phases $\Out$
and $\In$ with widths $L_\Out$ and $L_\In$, respectively. We take the
layers parallel to the $xy$ plane and choose the unit cell $0<z<L =
L_\Out+L_\In$ such that the region $0<z<L_\Out$ corresponds to phase
$\Out$ and $L_\Out<z<L$ to phase $\In$. Obviously the protons can
entrain the neutrons only in $z$ direction. In the rest frame of the
superfluid neutrons, the solution for the velocity potential reads
\begin{equation}
  \phi(\rv) =
  \begin{dcases}
    -\frac{1-\gamma}{L_\Out/L_\In+\gamma}\, z u_{p,z} &
        \text{for~} 0\leq z \leq L_\Out\,, \\
    \frac{1-\gamma}{1+\gamma L_\In/L_\Out}\,(z-L) u_{p,z} &
        \text{for~} L_\Out \leq z \leq L \,.
  \end{dcases}
  \label{phi-lasagne}
\end{equation}
From this solution, one can readily obtain the density of bound
neutrons (more precisely, the $zz$ component of the matrix
$\nnmbound$; all other components vanish):
\begin{equation}
  \nnboundel{zz} = \nnavg \frac{(1-\gamma)^2 L_\Out L_\In}
     {(L_\Out+\gamma L_\In)(L_\In+\gamma L_\Out)}\,.
\label{nbound-lasagne}
\end{equation}
In practice, $\nnboundel{zz}$ is much smaller than $\nnavg$
($\nnboundel{zz}/\nnavg\lesssim 0.03$) because the plates are only
found in the deepest layers of the neutron-star crust
\cite{MartinUrban2015}, where the density of the gas is quite large
($\gamma \gtrsim 0.7$).

\subsection{Numerical solution}
\label{sec:numeric}
In 2D and 3D, the situation is more difficult if one considers instead
of an isolated cluster a periodic lattice of clusters. Because of the
different geometries of the clusters and of the lattice, the solution
of the Laplace equation together with the boundary condition can only
be obtained numerically in this case.

\begin{figure}
  \includegraphics[width=5cm]{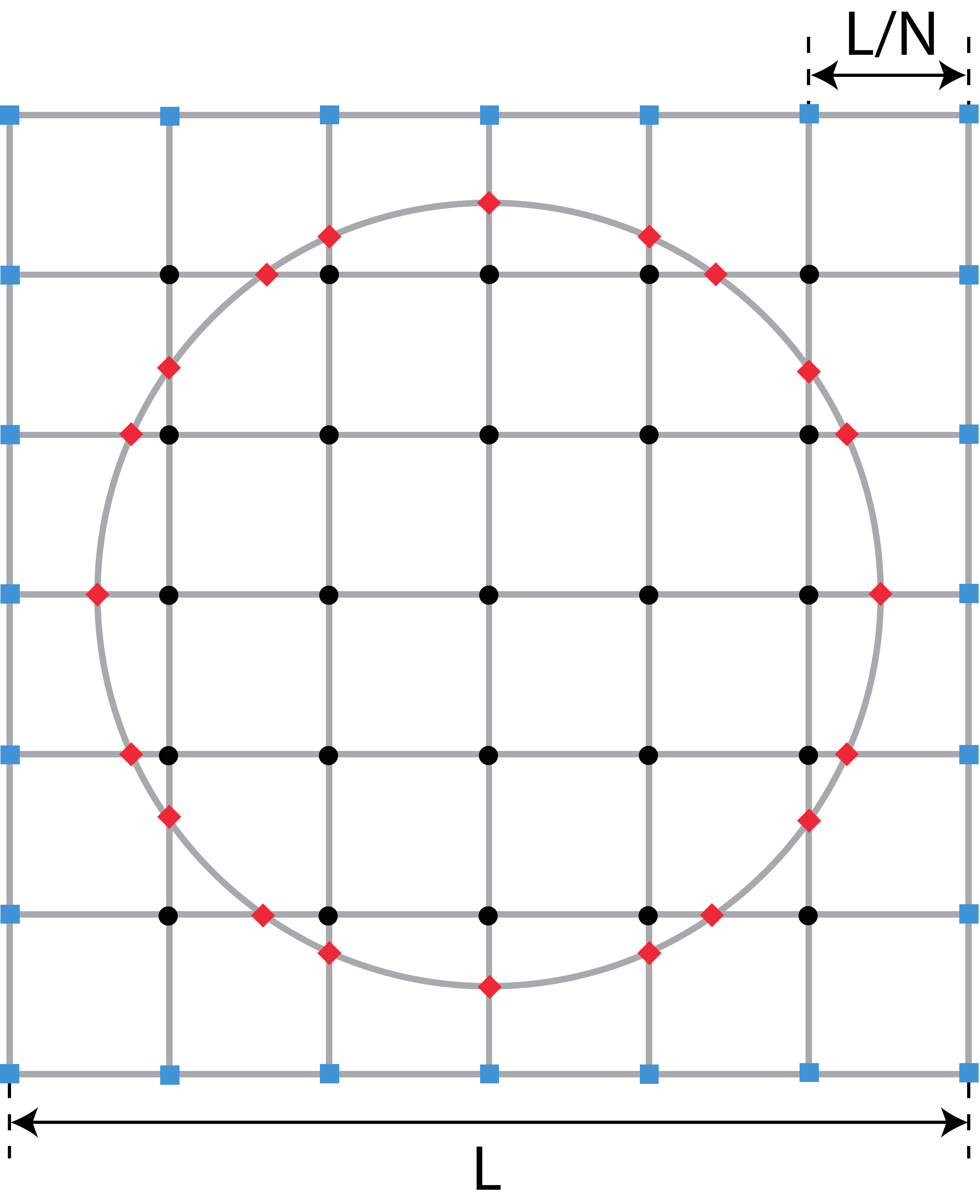}
  \caption{(Color online) Schematic illustration of the discretization
    of a simple cubic cell with a spherical cluster in its
    center.}
  \label{fig:mesh}
\end{figure}

We start by discretizing the cell space with a regular mesh of $N$
points per row. Note that if the unit cell is not cubic (as in the
hexagonal 2D case, see \Sec{sec:geometries}), the rows are not
orthogonal to one another. The cluster surface is approximated by a
set of $N_S$ points given by the intersections of the mesh lines with
the cluster surface. As an example, \Fig{fig:mesh}
illustrates the case of a spherical cluster in a simple cubic
cell. Points belonging to the cell mesh are shown as black circles and
blue squares, those belonging to the cluster surface as red
diamonds. Because of periodicity, points lying on opposite edges of
the cell, shown as blue squares, are equivalent to each other. In
total, the number of independent points is $\mathcal{N} =
N^D+N_S$. The numerical method for treating the periodicity is well
described in \Ref{EvansOkolie1982}.

Due to the space discretization, the differential equation
(\ref{eq:laplace-general})--(\ref{eq:boundary-condition3}) can be
written as a linear system of equations. The solution is represented
as a vector $\phiv$ of dimension $\mathcal{N}$ that contains the
values of $\phi(\xv_i)$, i.e., the solution of the differential
equation in the points $\xv_i$. In matrix form, the linear system of
equations is written as
\begin{equation}
  \mat{C} \phiv = \yv \, .
  \label{eq:laplace-linear}
\end{equation}
The elements of the $\mathcal{N}\times\mathcal{N}$ matrix $\mat{C}$
are the coefficients of the $\phi(\xv_i)$ in the discretized versions
of the Laplace equation (\ref{eq:laplace-general}) for all but one
mesh points and of the boundary condition
(\ref{eq:boundary-condition2}) for the $N_S$ surface points. To obtain
a closed system, the Laplace equation in one of the mesh points, say,
$\xv_{i_0}$ (we choose it to be the center of the cell), is replaced
by $\phi(\xv_{i_0}) = 0$, since otherwise $\phi$ would only be
determined up to an additive constant. The vector $\yv$ of dimension
$\mathcal{N}$ on the right-hand side of \Eq{eq:laplace-linear}
contains the inhomogeneities arising from the boundary condition
(\ref{eq:boundary-condition2}) due to the non-vanishing value of
$\uv_p$.
Explicitly, its components read as
\begin{align} y_{i\notin \Omega} =& 0\,,\\ y_{i\in \Omega} =&
  (n_{n,\Out} - n_{n,\In})\, \Sv_i \cdot \vb{u}_p \,,
\end{align}
where $\Omega$ denotes the surface points and $\Sv_i$ is the
normal vector in the surface point $i$.

Let us also be more specific concerning the calculation of the
matrix $\mat{C}$. The rows $i\neq i_0$ of the matrix $\mat{C}$ are
defined as follows:
\begin{align}
  (\mat{C}\phiv)_{i\notin\Omega} =& \laplace \phi(\xv_i)\,,
    \label{eq:cphi_inotins}\\
  (\mat{C}\phiv)_{i\in \Omega} =&
      \Sv_i \cdot \left(n_{n,\Out} \nablav \phi_{\Out}(\xv_i) 
     - n_{n,\In} \nablav \phi_{\In}(\xv_i)\right)\,,
  \label{eq:cphi_iins}
\end{align}
while the $i_0$-th row simply reads $C_{i_0j} = \delta_{i_0j}$. The
Laplacian in \Eq{eq:cphi_inotins} is expressed in terms of the second
partial derivatives that
are obtained by inverting the Taylor expansion
\begin{multline}
  \phi(\xv_j) = \phi(\xv_i) +
  \sum_{\mu = 1}^D \left. \frac{\partial \phi(\xv)}{\partial x_\mu} \right
  |_{\xv_i} ( x_{j,\mu} - x_{i,\mu} ) \\
  + \frac{1}{2} \sum_{\mu, \nu = 1}^D \left. \frac{\partial^2
  \phi(\xv)}{\partial x_\mu \partial x_\nu} \right
  |_{\xv_i} ( x_{j,\mu} - x_{i,\mu} ) ( x_{j,\nu} - x_{i,\nu} )\,,
\end{multline}
for $\{\xv_j\}$ the nine (in 3D) or five (in 2D) closest and linearly
independent points around $\xv_i$. The indices $\mu$ and $\nu$
correspond to the spatial directions. In the special case of a 2D mesh
with orthogonal axes (as in \Fig{fig:mesh}), one recovers in this way
exactly the expressions given in \Ref{Greenspan1964} for the
derivatives.
For the one-sided normal derivatives on the surface in
\Eq{eq:cphi_iins}, two different sets of points $\{\xv_j\}$ are used,
containing only surface points and points outside the cluster for
$\nablav\phi_{\Out}$, and only surface points and points inside the
cluster for $\nablav\phi_{\In}$.

In order to reduce the size of the matrix $\mat{C}$ in memory, we use
a sparse matrix storage (i.e., only non-zero matrix elements are
stored). Unfortunately, the solution of \Eq{eq:laplace-linear} cannot
be found with iterative methods (e.g., Gauss-Seidel) because the
matrix is not positive definite. So a direct LU decomposition is
needed, during which the size of the matrix blows up, which limits the
maximum size of $N$.
\section{Geometries}
\label{sec:geometries}
\subsection{Body-Centered Cubic lattice (3D)}
\label{sec:bcc}
\begin{figure}
  \includegraphics[width=5cm]{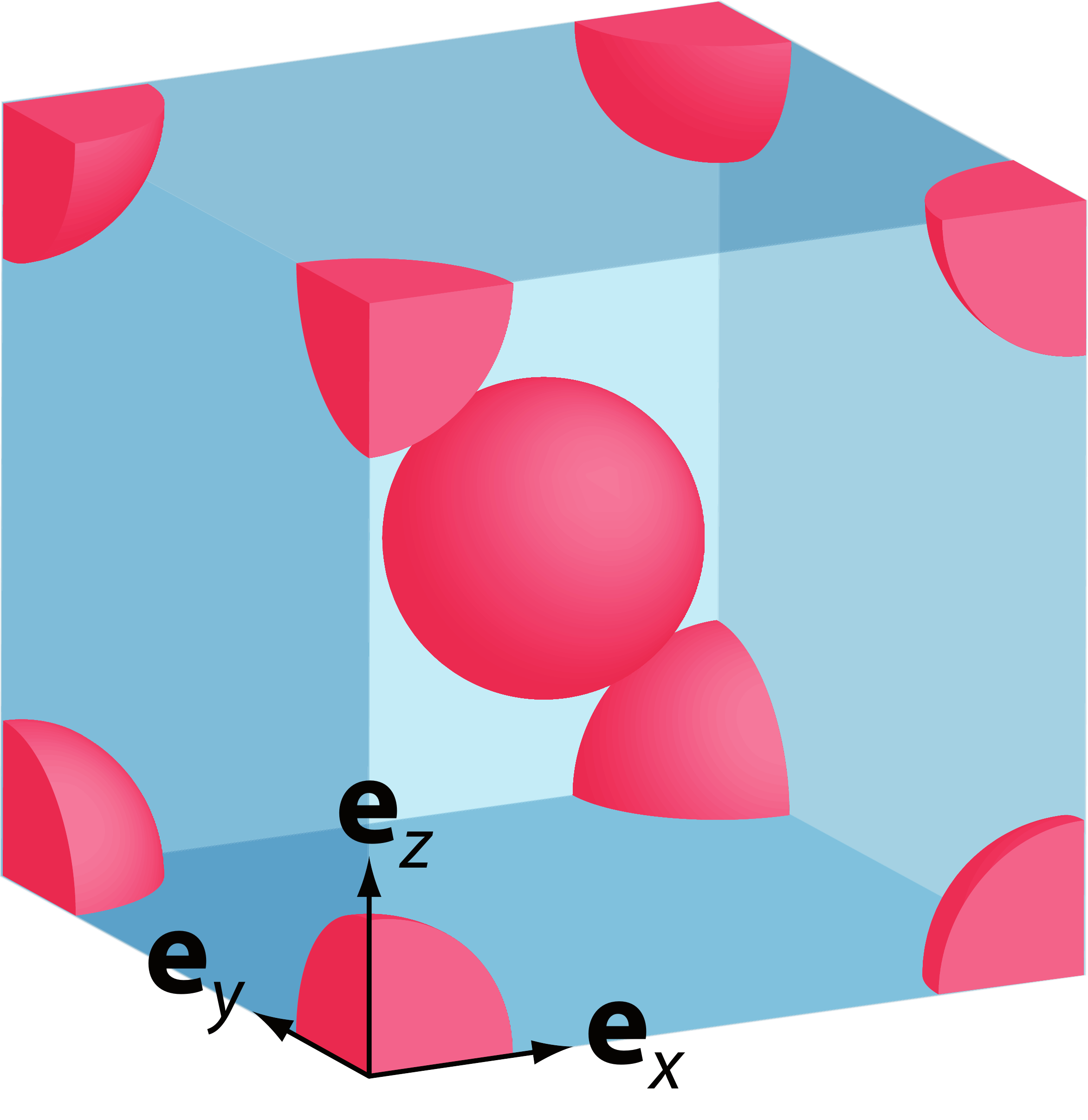}
  \caption{(Color online) Primitive cell of a BCC lattice of spherical
    clusters.}
  \label{fig:bcc}
\end{figure}

In the less dense parts of the inner crust, one expects a Coulomb
lattice of spherical clusters. The most favorable arrangement in space
is probably a BCC lattice \cite{Oyamatsu1984}. The
primitive cell of this lattice, \Fig{fig:bcc},
has one cluster at its center and one eighth at each corner, i.e., it
contains in total two clusters.

The BCC primitive cell presents symmetries simplifying the expressions
for the average current and the kinetic energy. Assuming a velocity
$\uv_p$ in direction $x$ and $\uv_n = 0$, the average neutron current
reads
\begin{equation}
  \jvb_n = \nnmbound \uv_p =
  \begin{pmatrix}
    \nnboundel{11} \\
    \nnboundel{21} \\
    \nnboundel{31}
  \end{pmatrix}
  u_p
  \, ,
\end{equation}
with $\nnboundel{ij}$ the elements of the matrix $\nnmbound$ in the
basis $\{\ev_x,\ev_y,\ev_z\}$. Because of the symmetry $y
\leftrightarrow -y$ and $z \leftrightarrow -z$, the current $\jvb_n$
cannot have any component in $y$ or $z$ directions, i.e., the
off-diagonal elements $\nnboundel{21}$ and $\nnboundel{31}$ must
vanish. Repeating the same arguments for velocities $\uv_p$ in $y$ or
$z$ directions, one finds that all off-diagonal elements are zero.

Furthermore, the directions $x$, $y$ and $z$ are equivalent in BCC
symmetry. Thus all diagonal terms are equal, and the matrix simply
reduces to a scalar matrix $\nnmbound = \nnbound \UnitM{3}$. So
finally, in the BCC lattice, for $\uv_n = 0$, $\jvb_n$ and $\ekinn$
are simply given by
\begin{equation}
  \jvb_n = \nnbound \uv_p \quad \text{and}\quad \ekinn = \frac{m}{2}
  \nnbound \uv_p^2 \, ,
\end{equation}
and there is no effect of anisotropy.
\subsection{Hexagonal lattice (2D)}
\label{sec:hexagonal}
\begin{figure}
  \includegraphics[width=5cm]{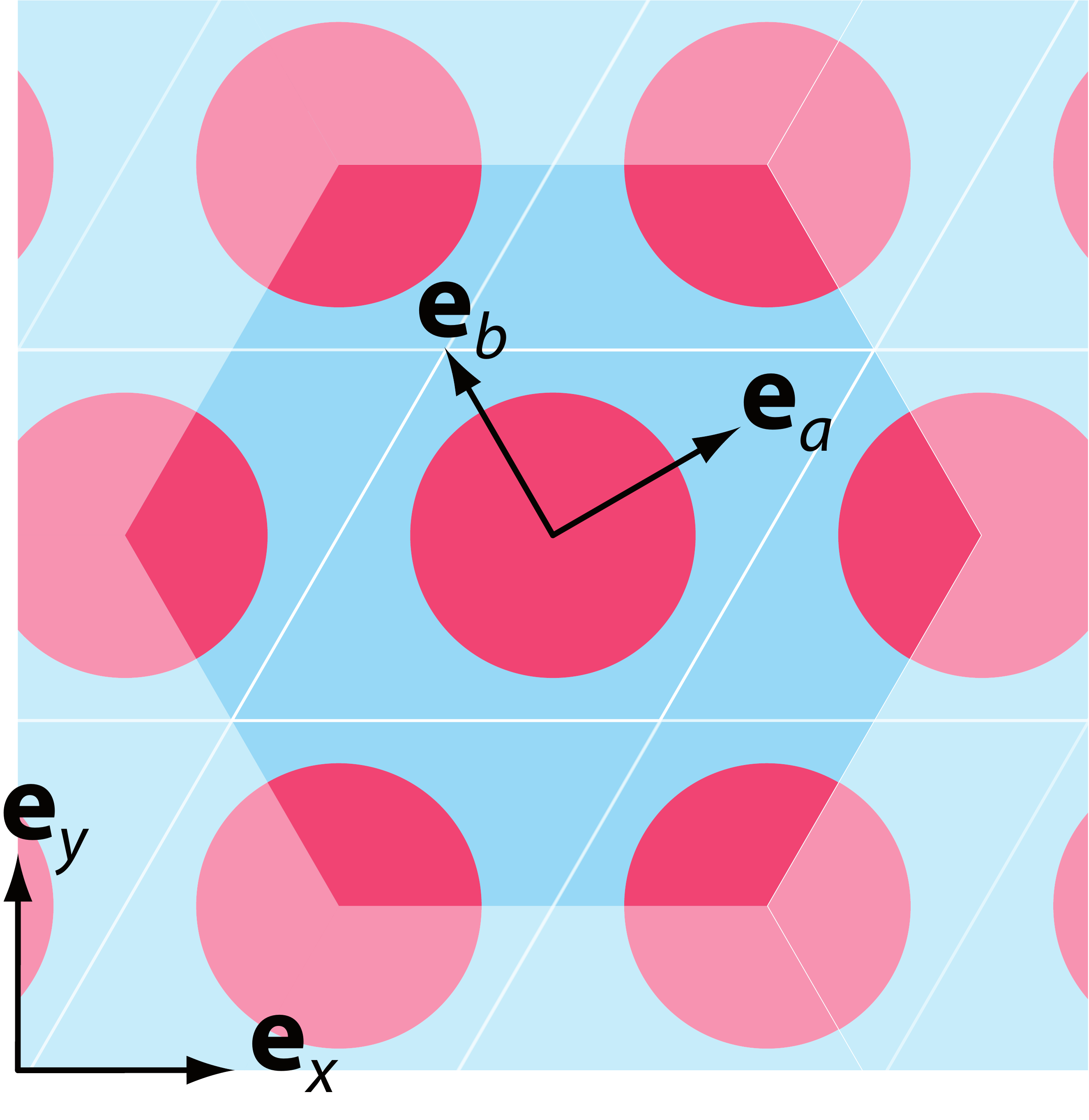}
  \caption{(Color online) Cut through a hexagonal lattice of
    cylindrical rods. The primitive cell is the parallelogram
    delimited by the white lines.}
  \label{fig:hexa}
\end{figure}

Deeper in the crust, clusters are supposed to be rods of bound
nucleons \cite{Ravenhall1983}. In this case the most favored
arrangement with respect to the Coulomb energy is a hexagonal lattice
\cite{Oyamatsu1984}. The primitive cell is a rhombus of
side length $L$, height $\sqrt{3}L/2$ and an angle of $\pi/3$, as
shown in Fig.~\ref{fig:hexa}.
From the symmetry of the cell it is clear that the eigenvectors of
$\nnmbound$ are $\ev_a$, $\ev_b$ and $\ev_z$ with:
\begin{equation}
  \begin{pmatrix}
    \ev_a \\ \ev_b
  \end{pmatrix}
  =
  \begin{pmatrix}
    \nicefrac{\sqrt{3}}{2} & \nicefrac{1}{2} \\
   -\nicefrac{1}{2} & \nicefrac{\sqrt{3}}{2}
  \end{pmatrix}
  \begin{pmatrix}
    \ev_x \\ \ev_y
  \end{pmatrix}
  \, .
\end{equation}
The vectors $\ev_a$ and $\ev_b$ are shown in \Fig{fig:hexa}. The three
directions $(a,b,z)$ are, however, not equivalent, thus in the basis
$\{\ev_a, \ev_b, \ev_z\}$ the diagonal elements (eigenvalues) of
$\nnmbound$ are all different: $\nnboundel{11} \ne \nnboundel{22} \ne
\nnboundel{33}$. Let us note that the rods are invariant with respect
to the $z$ axis, i.e., all neutrons can move freely in that direction,
consequently $\nnboundel{33} = 0$.
\section{Results}
\label{sec:results}
\subsection{Microscopic flow}
\label{sec:micro-flow}
We solve \Eqs{eq:laplace-general}--(\ref{eq:boundary-condition2}) for
a fixed velocity $\uv_p$ of the clusters. As input for the radius of
the clusters, the densities inside and outside the clusters, and the
cell size, we use results obtained in \Ref{MartinUrban2015} within the
Extended Thomas-Fermi (ETF) method with a Skyrme energy-density
functional (SLy4).

\begin{figure*}
  \subfigure[~Velocity field ($z=0$)]
  {\includegraphics[height=6cm]{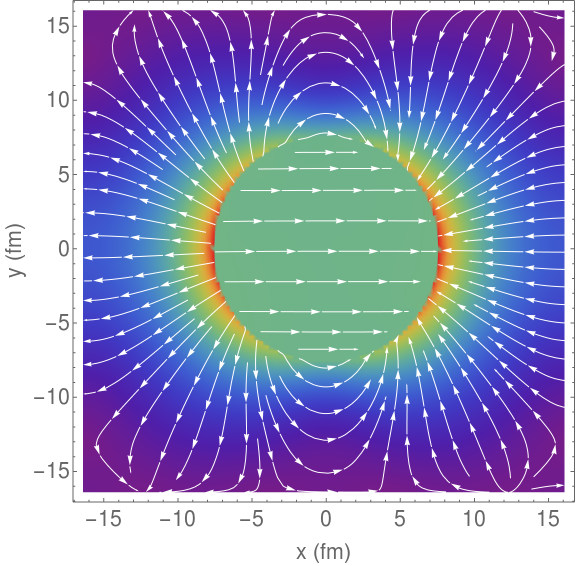}
  \label{fig:stream_bcc_center}} \hspace{1.5cm}
  \subfigure[~Velocity potential ($z=0$)]
  {\includegraphics[height=5cm]{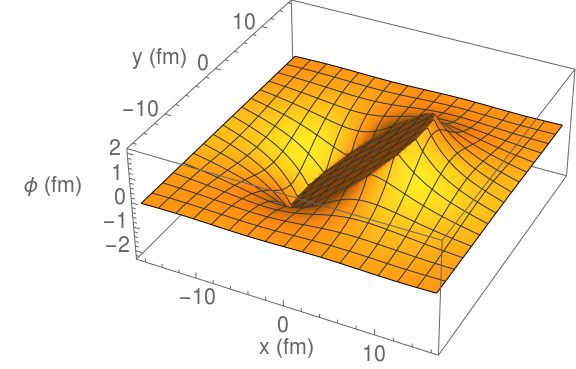}
  \label{fig:potential_bcc_center}}
  \subfigure[~Velocity field ($z=L/4$)]
  {\includegraphics[height=6cm]{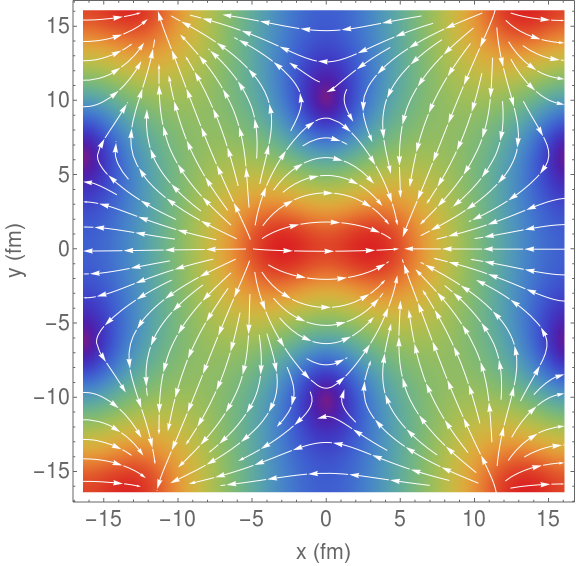}
  \label{fig:stream_bcc_middle}} \hspace{1.5cm}
  \subfigure[~Velocity potential ($z=L/4$)]
  {\includegraphics[height=5cm]{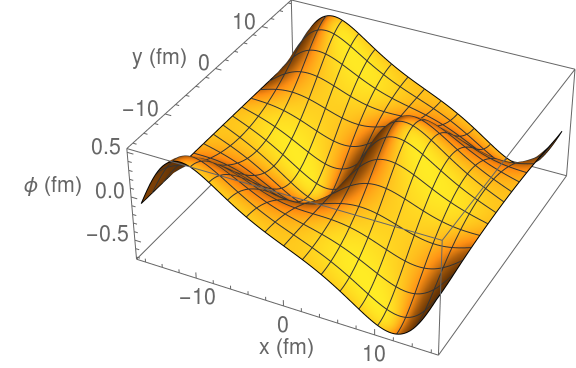}
  \label{fig:potential_bcc_middle}}
  \caption{(Color online) Streamlines and neutron speed (left) and
    velocity potentials (right) in a BCC cell of size $L=32.8$ fm,
    with a cluster of radius $R=7.54$ fm moving with velocity $\uv_p =
    \ev_x$. The neutron density inside the cluster is $n_{n,\In} =
    0.0973$ fm$^{-3}$ and outside $n_{n,\Out} = 0.0412$ fm$^{-3}$ (the
    cluster and cell properties were obtained from calculations
    described in \Ref{MartinUrban2015} and correspond to a baryon
    density of $n_B = 0.0485$ fm$^{-3}$).  In the left panels, the
    streamlines are shown as the white arrows, and the speed of the
    flow is indicated by the background color from dark purple
    (slowest) to red (fastest).}
  \label{fig:stream_bcc}
\end{figure*}

Figure~\ref{fig:stream_bcc}
shows streamlines (left panels) and velocity potential (right panels)
in a BCC cell, in the case of $\uv_p$ in $x$ direction. The
neutron-fluid streamlines are displayed as white arrows, they
characterize the flow direction and are tangential to the velocity
field vectors. The background color scheme indicates the speed, from
dark purple in the slowest zones to red in the fastest ones. We chose
two cuts through the cell parallel to the $xy$ plane. The upper panels
correspond to the plane through the center of the cell ($z=0$), while
the lower panels correspond to a plane between the clusters
($z=L/4$). In \Fig{fig:stream_bcc_center} the neutron velocity inside
the cluster $\vv_{n,\In}$ is practically constant but lower than the
velocity $\uv_p$ of the surface. Here the ratio between the fluid and
the surface velocity is $|\vv_{n,\In}| / |\uv_p| = 0.284$, which can
be compared with the analytic result (\ref{phi-spherical}) for the
neutron velocity inside a cluster moving through an infinite neutron
gas: $(1-\gamma)/(1+2\gamma) = 0.315$

Furthermore one sees that neutrons between the clusters move in the
opposite direction. The velocity discontinuity at the cluster surface
satisfies the boundary condition (\ref{eq:boundary-condition2}) of the
conservation of the neutron current crossing the surface. When going
away from the cluster surface, we observe that the speed decreases
because the flux is spread over a larger surface.
Figures~\ref{fig:stream_bcc_middle}--(d) show the plane between the
clusters at $z = L/4$. One can observe on the edges of the cell the
periodicity of the field. The five red areas correspond to the regions
that are closest to the clusters.

\begin{figure*}
  \subfigure[~Velocity field ($\uv_p = \ev_a$)]
  {\includegraphics[height=4.5cm]{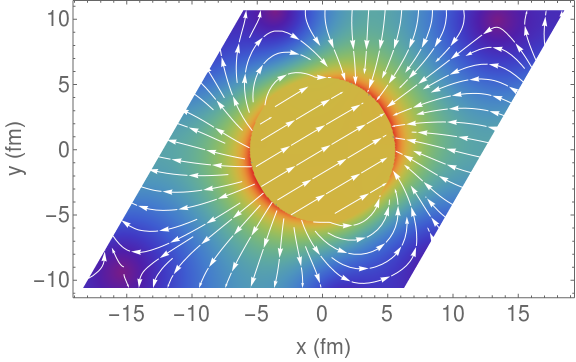}
  \label{fig:stream_hexa_e1}} \hspace{1.5cm}
  \subfigure[~Velocity potential ($\uv_p = \ev_a$)]
  {\includegraphics[height=4.5cm]{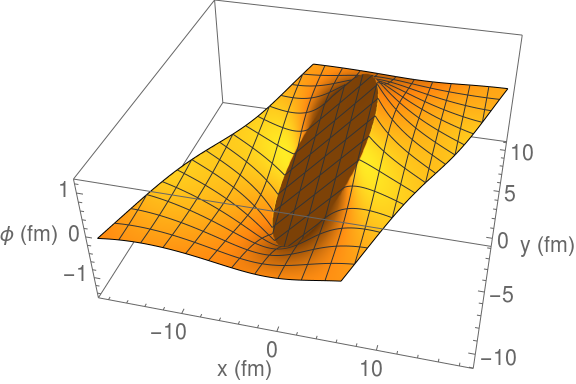}
  \label{fig:potential_hexa_e1}}\\
  \subfigure[~Velocity field ($\uv_p = \ev_b$)]
  {\includegraphics[height=4.5cm]{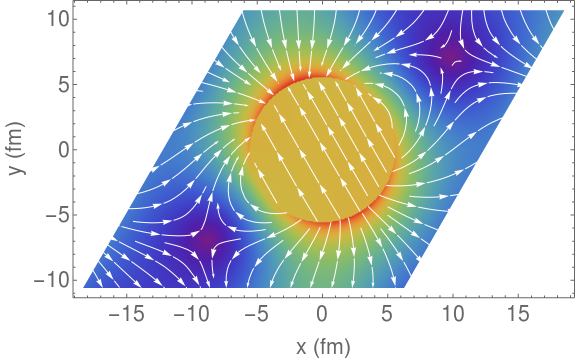}
  \label{fig:stream_hexa_e2}} \hspace{1.5cm}
  \subfigure[~Velocity potential ($\uv_p = \ev_b$)]
  {\includegraphics[height=4.5cm]{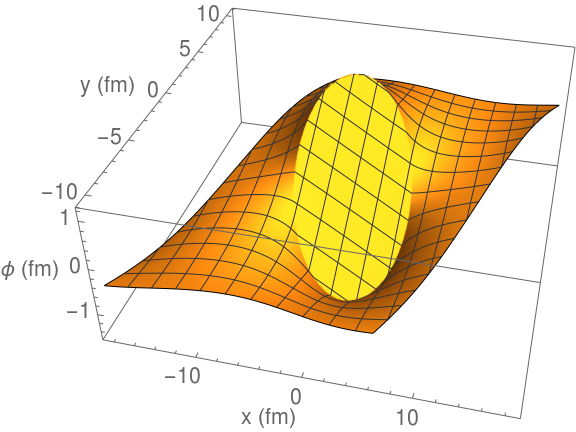}
  \label{fig:potential_hexa_e2}}
  \caption{(Color online) Same as \Fig{fig:stream_bcc}, but for a
    hexagonal cell of size $L= 24.7$ fm, containing a cylindrical rod
    of radius 5.53 fm moving with velocity $\uv_p=\ev_a$ (upper
    panels) or $\ev_b$ (lower panels). The neutron density inside the
    rod is 0.0942 fm$^{-3}$ and outside 0.0528 fm$^{-3}$
    (corresponding to a baryon density of $n_B = 0.0624$ fm$^{-3}$).}
  \label{fig:stream_hexa}
\end{figure*}

Let us now discuss the case of the hexagonal lattice shown
Fig.~\ref{fig:stream_hexa}.  Qualitatively, the behavior is similar to
the one observed in the BCC lattice. However, in contrast to the BCC
case, the hexagonal primitive cell is not isotropic. Thus we performed
calculations with velocities $\uv_p$ in the directions of the
eigenvectors $\ev_a$ and $\ev_b$ (cf. \Sec{sec:hexagonal}). One can
clearly see a strong difference of the periodic behavior between
Fig.~\ref{fig:stream_hexa_e1} and Fig.~\ref{fig:stream_hexa_e2},
especially at the corners of the primitive cell. In
Fig.~\ref{fig:stream_hexa_e2}, the streamlines continue straight to
the next cell, while in Fig.~\ref{fig:stream_hexa_e1} they deviate
from their initial trend $\ev_a$. Instead of exiting or entering
through the corners of the cell, the flow passes through its sides and
then through the neighboring clusters situated in the directions of
the translation vectors $\av_1$ and $\av_2$ of the Bravais lattice
(parallel to the white lines in Fig.~\ref{fig:hexa}). Hence, the
currents and the energies depend on the direction of $\uv_p$.
Nevertheless, the anisotropy effect on the ratio
$|\vv_{n,\In}|/|\uv_p|$ is very weak, numerically one finds $0.244$
and $0.248$ in the cases of $\uv_p$ in direction $\ev_a$ and $\ev_b$,
respectively. Similarly to the BCC case, this ratio is somewhat lower
than the analytical result \Eq{phi-spaghetti} for a single rod in an
infinite gas, $|\vv_{n,\In}|/|\uv_p| = 0.281$.

\subsection{Cluster effective mass and superfluid density}
\label{sec:effmass}
With the help of Eq.~(\ref{eq:jmean}), which is equivalent to
averaging the microscopic current over the cell, one obtains the
macroscopic quantities $\nnmsuper$ and $\nnmbound$. In
\Sec{sec:entrainment}, they were interpreted as if $\nnmsuper$ were
the neutrons that move independently of the clusters while $\nnmbound$
are the neutrons moving with the clusters. However, the preceding
discussion of the microscopic flow shows that this is a simplified
picture. In the BCC case, staying within this picture, we can define a
cluster effective mass number
\begin{equation}
\Aeff = \Neff + Z = \frac{1}{2} V_\cell \nnbound + Z\,,
\end{equation}
where the factor $1/2$ accounts for the fact that there are two
clusters per cell and $Z$ is the number of protons in each cluster.

The cluster effective mass plays an important role for the calculation
of the lattice phonons, as discussed, e.g., in
\cite{Sedrakian1996,Magierski2004,ChamelPage2013}. It can be compared
with the trivial result one obtains by counting all nucleons that are
geometrically located inside the cluster, $\Arcluster = \Nrcluster +
Z$.

\begin{figure}
  \includegraphics[width=5cm]{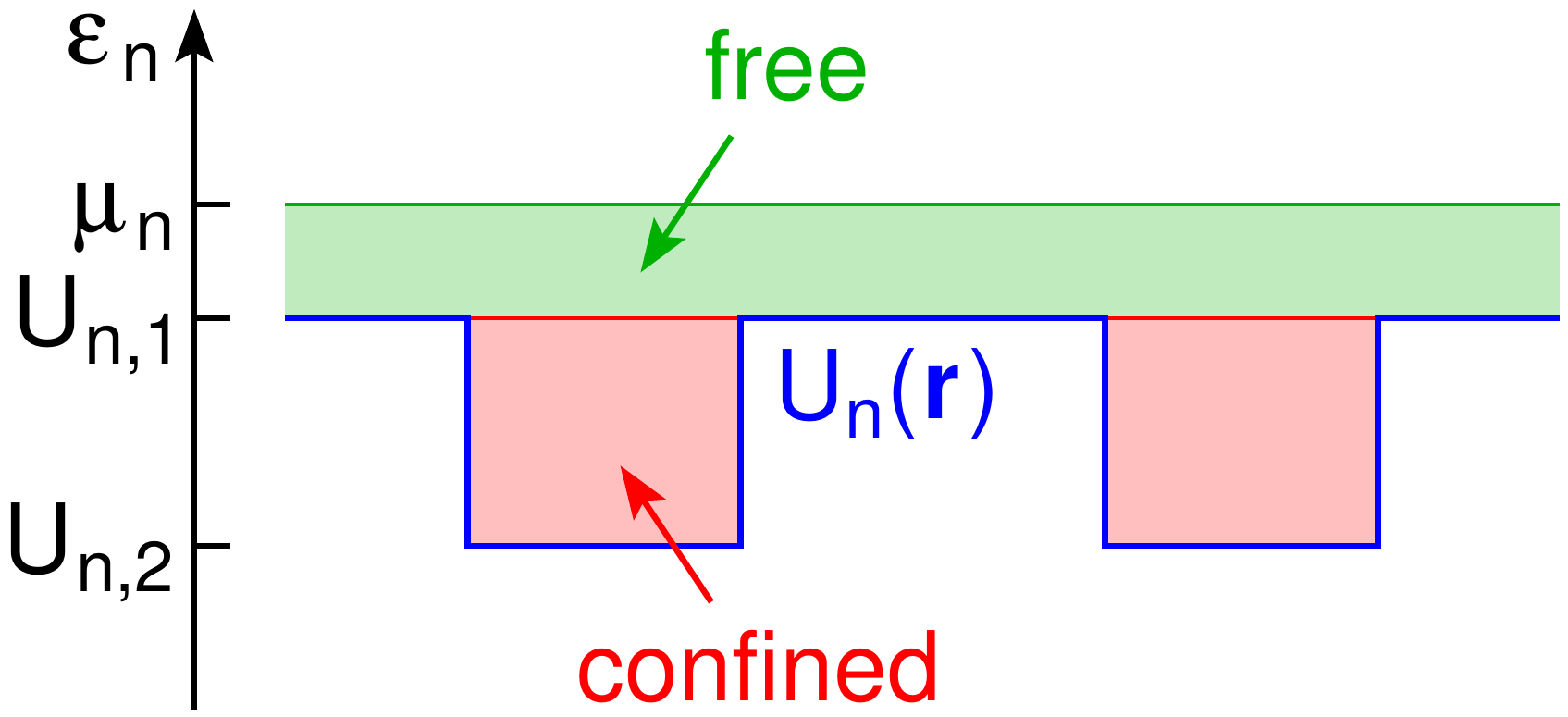}
  \caption{(Color online) Schematic illustration of the definition of
    free and confined neutrons in energy space.}
  \label{fig:ecluster}
\end{figure}

However, it might be more appropriate to define the cluster in
energy space (e-cluster \cite{Papakons2013}). In this picture,
neutrons are considered \emph{free} or \emph{confined}
\cite{ChamelCarter2006} (the word \emph{bound} is also employed
\cite{Papakons2013,CarterChamel2005} but should not be confused with
the effectively bound neutrons defined in \Sec{sec:entrainment})
depending on their energy and independently of their position, i.e.,
free neutrons may also be located inside the cluster. In our
approximation of constant densities in the two phases, the neutron
Hartree-Fock mean field $U_n(\rv)$ is also constant in each phase and
takes the values $U_{n,\Out}$ in the gas and $U_{n,\In}$ in the
cluster. Confined neutrons are characterized by a single-particle
energy $\epsilon_n(\kv) = k^2/(2m_n^*)+U_n$ that lies below the mean
field in the gas, $\epsilon_{n}(\kv) < U_{n,\Out}$, while the
single-particle energy of free neutrons lies above, $\epsilon_{n}(\kv)
> U_{n,\Out}$, see \Fig{fig:ecluster}.
Hence, the density of confined neutrons inside the cluster is in this
picture given by
\begin{equation}
  \nnconfinedin = \frac{1}{3\pi^2}[2m^*_{n,\In}(U_{n,\Out}-U_{n,\In})]^{3/2}\,,
\label{nconfined}
\end{equation}
with $m^*_{n,i}$ the neutron effective mass calculated in phase $i$,
and the remaining neutrons inside the cluster are free\footnote{Here
  we do not distinguish between localized and unlocalized unbound
  neutrons \cite{Papakons2013}.},
\begin{equation}
  \nnfreein = n_{n,\In} - \nnconfinedin\,.
\end{equation}
The effective neutron and mass numbers of the cluster (in energy
space) are therefore
\begin{equation}
\Necluster = \frac{4 \pi}{3} R^3 \nnconfinedin
\label{Necluster}
\end{equation}
and $\Aecluster = \Necluster + Z$. The mean fields $U_{n,i}$ and
effective masses $m^*_{n,i}$ in \Eq{nconfined} are calculated with the
same Skyrme functional (SLy4) that was used in the ETF calculation of
the cell properties \cite{MartinUrban2015}.

\begin{figure}
  \includegraphics[width=8cm]{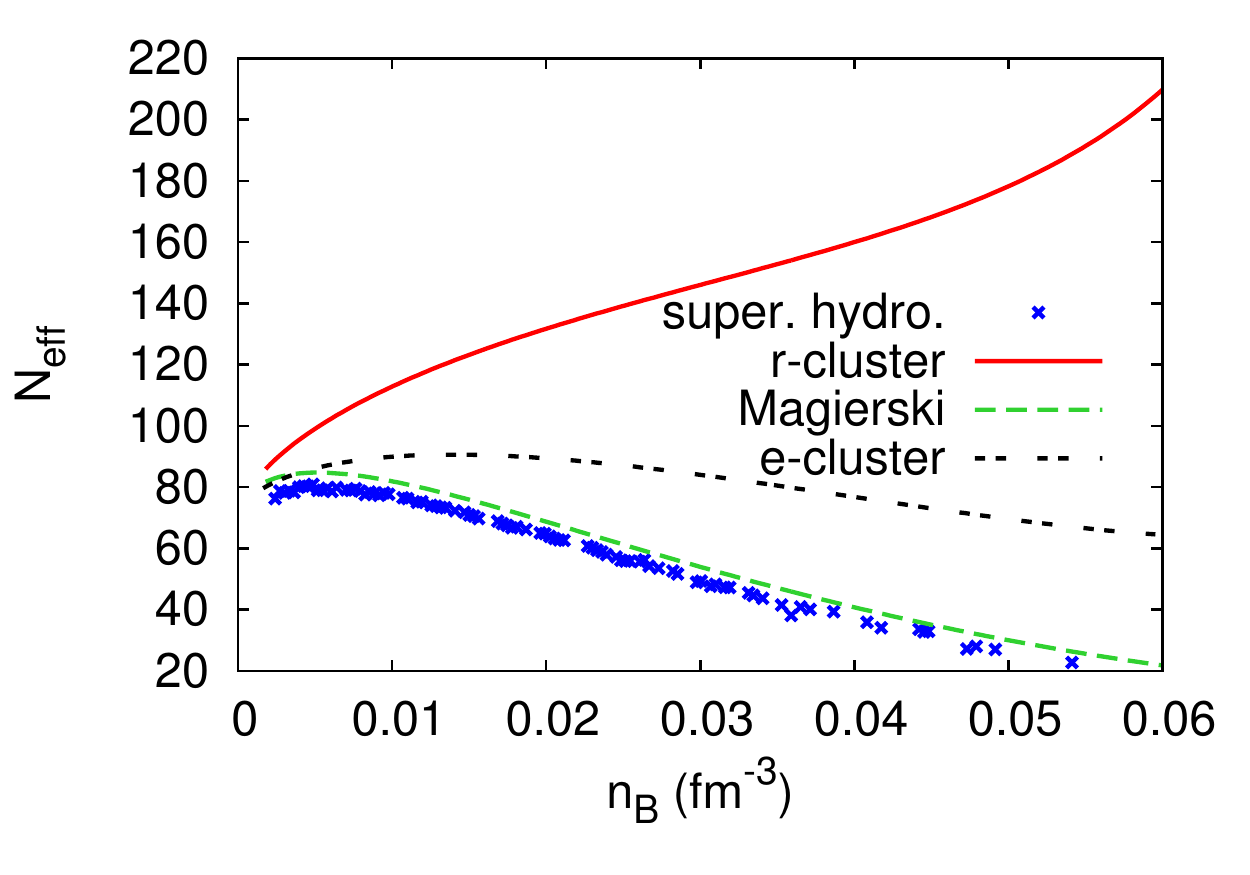}
  \caption{(Color online) Effective neutron number of the clusters
    moving through the neutron gas as a function of the baryon density
    $n_B$. The results of our numerical calculations (blue crosses)
    are compared with the result of \Eq{NeffMagierski} by Magierski
    and Bulgac
    \cite{MagierskiBulgac2004,MagierskiBulgac2004NPA,Magierski2004}
    for an isolated cluster in a uniform neutron gas (green
    dashed line), and with the neutron numbers
    (\ref{Nrcluster}) and (\ref{Necluster}) of the cluster defined in
    coordinate (red solid line) and energy (black double-dashed
      line) space, respectively.}
  \label{fig:effmass}
\end{figure}

In \Fig{fig:effmass}, 
we compare the effective neutron numbers of the clusters obtained
within the different approaches as functions of the baryon density
$n_B = \nnavg+\npavg$. At low density, i.e., close to the outer crust,
the density of the neutron gas is very low and all approaches converge
towards the same result. However, at higher density, when the density
of the neutron gas increases, the approaches start to differ
considerably. More and more neutrons that are located inside the
clusters (in coordinate space) are not bound in energy
space. Therefore, the number of neutrons in the e-cluster (black
double-dashed line) is considerably smaller than the number of
neutrons in the r-cluster (red solid line).

However, the effective neutron number obtained within the present
superfluid hydrodynamics approach (blue crosses) is even smaller: at
the highest densities where one still expects the BCC lattice, one
finds $\Necluster/\Nrcluster\approx 0.3$, while superfluid hydrodynamics
predicts $\Neff/\Nrcluster\approx 0.1$. Quite surprisingly, even at the
highest densities, where the unit cell is not very large compared to
the cluster size, our numerical results stay quite close to the
analytical ones, \Eq{NeffMagierski} one would obtain for an isolated
cluster (green dashed line).

Concerning the (small) difference between the numerical results and
those of \Eq{NeffMagierski}, one might think that it comes from the
restriction of the integration to a finite volume. Actually, one can
easily derive a modified version of \Eq{NeffMagierski} where one
integrates the neutron current $n_n \nablav \phi$ only up to the
WS radius instead of infinity, but it turns out that the
difference is negligible. The main reason for the difference between
the numerical results and those of \Eq{NeffMagierski} is the change of
the velocity potential $\phi$ itself due to the periodic boundary
conditions.

\begin{figure}
  \includegraphics[width=8cm]{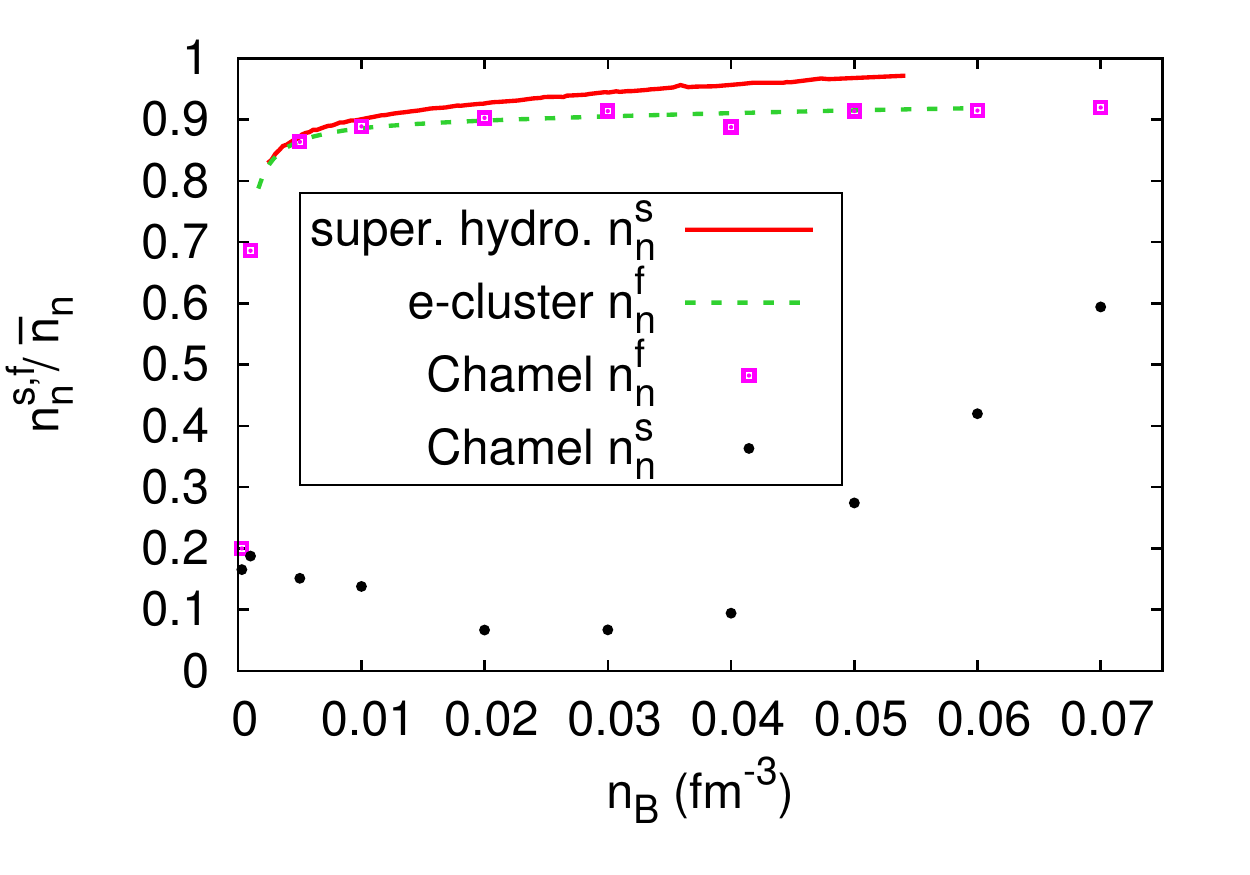}
  \caption{(Color online) Fraction of superfluid neutrons,
    $\nnsuper/\nnavg$ as a function of the baryon density
    $n_B$. Results of the present superfluid hydrodynamics approach
    (red solid line) are compared with the result of band-structure
    calculations by Chamel \cite{Chamel2012} (black circles). We
    display also our results for the fraction of (energetically) free
    neutrons $\nnfree/\nnavg$ (green dashes) and those obtained within
    the band-structure approach \cite{Chamel2012} (purple squares).
  \label{fig:nsnn-ratio}}
\end{figure}

Another quantity of interest is the superfluid density $\nnsuper$. In
\Fig{fig:nsnn-ratio}
we show the superfluid fraction $\nnsuper/\nnavg$ as a function of the
baryon density $n_B$. Unfortunately, we cannot perform numerical
calculations at very low total densities (as they prevail near the
outer crust), because the unit cells become too large. But it seems
that at these low densities, the superfluid density obtained within
our hydrodynamic approach (solid red line) agrees approximately with
the density of free neutrons (green dashed line). At higher total
neutron densities, the superfluid fraction is larger than the density
of free neutrons and it increases rapidly above 90\,\%, exceeding
$97$\,\% at the transition towards the 2D phase.

We compare these results with those obtained by Chamel
\cite{Chamel2012} in the framework of the band theory for neutrons
(black circles). This theory is analogous to the band theory in
solid-state physics to describe electrons in the periodic Coulomb
potential of a crystal \cite{AshcroftMermin}. In the inner crust of a
neutron star, one has instead neutrons in the periodic mean field
generated by the clusters. The superfluid density is in this approach
obtained from an average of the Fermi velocity over the (highly
nontrivial) Fermi surface
\cite{CarterChamel2005,Chamel2006,ChamelPage2013}. While in our
hydrodynamic approach the superfluid density is higher than the
density of free neutrons, the band-structure calculation predicts a
much lower superfluid density. Possible reasons for this discrepancy
will be discussed in \Sec{sec:discussion}.

As a consistency check, we also compare our results for the fraction
of free neutrons with those of the band-structure approach (purple
squares), and for this quantity the agreement is excellent in spite of
the crude approximations (sharp interface between the cluster and the
gas, Thomas-Fermi approximation for the density of states) underlying
\Eq{nconfined}.

\begin{figure}
  \includegraphics[width=8cm]{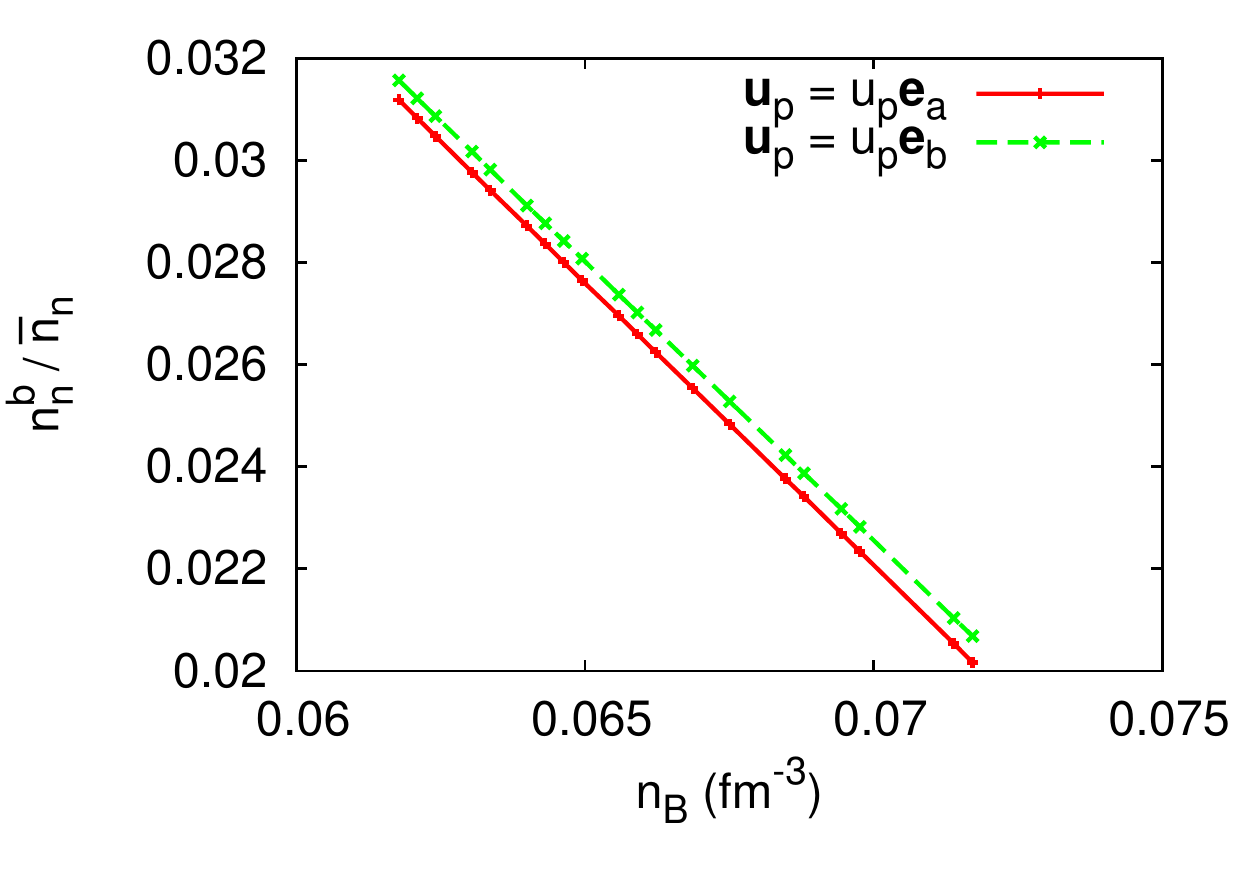}
  \caption{(Color online) Effective densities of bound neutrons in the
    2D (spaghetti) phase for velocities in the directions of the two
    eigenvectors $\ev_a$ and $\ev_b$ as functions of total baryonic
    density.}
  \label{fig:aniso_hexa}
\end{figure}

So far we discussed the BCC lattice, where the densities of bound and
superfluid neutrons are scalar quantities. The situation is different
in the 2D hexagonal lattice of rods. In this case, if the velocity is
parallel to the rods ($z$ direction), the neutrons can move
independently of the protons and the superfluid fraction is
100\,\%. In the transverse plane, however, there is some
entrainment. In \Fig{fig:aniso_hexa},
we show the densities of bound neutrons, $\nnbound$, for velocities in
the directions of the eigenvectors $\ev_a$ (red solid line) and
$\ev_b$ (green dashed line), as functions of the average neutron
density in the density range where we expect to find the 2D phase,
i.e., between $\sim 0.06$ and $0.07$ fm$^{-3}$
\cite{MartinUrban2015}. It can be seen that the anisotropy in the
transverse plane, i.e., the difference between the directions $a$ and
$b$, is very small.

\subsection{Application to glitches}
\label{sec:glitches}
Glitches correspond to a sudden transfer of angular momentum from the
superfluid to the normal parts of the star \cite{ShapiroTeukolsky}.
In the preceding sections, we discussed the densities of bound and
superfluid neutrons in the inner crust. These quantities play a
crucial role in the understanding of glitches in the neutron star
\cite{ChamelCarter2006,Chamel2013PRL}. In particular, as pointed out
in \Refs{AnderssonGlampedakis2012,Chamel2013PRL}, the
observed glitches of the Vela pulsar can hardly be understood with the
low superfluid fraction obtained in band structure theory. Since our
results for the superfluid fraction are very different from those of
band structure theory, let us discuss how this changes the conclusions
from the glitch data. In this subsection, we follow to a large extent
the arguments given in \Refs{ChamelCarter2006,Chamel2013PRL}.

Let us assume that the superfluid and the normal parts of the star
rotate at slightly different but spatially constant frequencies
$\Omega_s$ and $\Omega_b$, i.e., the velocity fields are given by
$\uv_p = \Omegav_b\times\rv$ and $\uv_n = \Omegav_s\times\rv$.

Note that $\uv_n$ has to be understood as the average velocity field
on length scales that are large compared to the distance between the
quantized vortices \cite{LandauLifshitz9}. If we consider, e.g., a
frequency of $\Omega_s = 100$ s$^{-1}$, the number of vortices per
area is \cite{PinesAlpar1985} $2m\Omega_s/(\pi\hbar) \approx
10^9$ m$^{-2}$, i.e., the vortices are separated by $\sim 30$
$\mu$m. Since this distance is many orders of magnitude larger than
the crystalline structures in the inner crust, one may use the results
for $\nnsuper$ and $\nnbound$ calculated for a uniform velocity field.

The total angular momentum of the star can now be decomposed into two
contributions,
\begin{equation}
  J = J_s + J_b = I_s \Omega_s + I_b \Omega_b \,,
  \label{eq:j_i}
\end{equation}
where $I_s$ and $I_b$ are the moments of inertia of the superfluid and
normal-fluid components, respectively
\footnote{Note that, unlike in \Ref{ChamelCarter2006}, there are no
  non-diagonal contributions to the angular momentum (contributions of
  $\Omega_s$ to $J_b$ and vice versa) because we are working in the
  chemical basis of superfluid and bound neutrons,
  cf. \Eq{ekindiagonal}.}:
\begin{equation}
  I_s = \int m \nnsuper r^2_\perp d^3r \,,\qquad
  I_b = \int m (\nnbound + n_p) r^2_\perp d^3r\,,
\end{equation}
with $r_\perp = r \sin\theta $ the radial distance from the rotation
axis.

As argued in \Refs{LinkEpstein1999,ChamelCarter2006}, the entire core
is probably rotating together with the non-superfluid part. Therefore,
the superfluid contribution comes only from the superfluid neutrons in
the inner crust, and the neutrons in the core are counted in
$\nnbound$, although they are of course not bound to clusters.

Between two glitches, the observable frequency $\Omega_b$ is slowly
decreasing because the emission of radiation leads to some loss of
angular momentum of the normal component. Let us denote by
$\Delta\Omega_b < 0$ the frequency change during the interglitch time.
The superfluid component, however, is supposed to slow down much less
than the normal component, e.g., because the vortices are
pinned. Hence, the superfluid component can serve as a reservoir of
angular momentum for the next glitch \cite{LinkEpstein1999}. A glitch
is interpreted as a sudden transfer of angular momentum from the
superfluid to the normal fluid component. However, during the short
duration of the glitch, the total angular momentum is
conserved. Therefore, the differences of the frequencies before and
after the glitch, denoted by $\delta\Omega_s$ and $\delta\Omega_b$,
satisfy
\begin{equation}
I_s \delta\Omega_s+I_b \delta\Omega_b = 0\,.
\label{angular-momentum-conservation}
\end{equation}
Since $\Omega_s-\Omega_b$ cannot become too large, $\Omega_s$ must in
average (after many glitches) decrease by the same amount as
$\Omega_b$, i.e.,
\begin{equation}
\langle \delta \Omega_s \rangle \geq \langle \Delta \Omega_b \rangle +
\langle \delta \Omega_b \rangle\,,
\label{spin-down}
\end{equation}
where the equality corresponds to the limiting case that the
superfluid does not slow down at all between two glitches
($\Delta\Omega_s = 0$). Combining \Eq{angular-momentum-conservation}
and (\ref{spin-down}), one arrives at the simple relation:
\begin{equation}
  \frac{I_s}{I} \geq -\frac{\langle \delta \Omega_b \rangle}{\langle
    \Delta \Omega_b \rangle} \equiv \mathcal{G} \,,
  \label{eq:isi_condition}
\end{equation}
with $I = I_s+I_b$ the total moment of inertia of the neutron star,
and $\mathcal{G}$ the \emph{coupling parameter}, which is closely
related to the \emph{pulsar activity parameter}
\cite{LinkEpstein1999}.

Following \Ref{Chamel2013PRL}, one can make some additional
approximations in order to obtain a quick estimate for the ratio
$I_s/I$. First, we write $I_s/I = (\Icrust/I)(I_s/\Icrust)$, where
$\Icrust$ is the moment of inertia of the crust. For the crustal
fraction of the moment of inertia, $\Icrust/I$, Lattimer and Prakash
\cite{LattimerPrakash2000} gave an approximate expression that depends
only on the pressure $\Pcore$ and density $\ncore$ at the crust-core
transition and on the total radius $R$ and mass $M$ of the star, but
does not require detailed knowledge of the high-density equation of
state (EOS) in the core. Moreover, making use of the thin crust
approximation \cite{Lorenz1993}, one can derive the following simple
expression for the superfluid contribution of the crustal moment of
inertia \cite{Chamel2013PRL}:
\begin{equation}
  \frac{I_s}{\Icrust} = \frac{1}{\Pcore} \int_{\Pdrip}^{\Pcore}
  \frac{\nnsuper}{n_B} \, dP \,,
  \label{eq:IsIcrust}
\end{equation}
where $\Pdrip$ is the pressure at the transition between the outer and
the inner crust. Here, we use the EOS of the ETF model of the inner
crust \cite{MartinUrban2015}. With our results for the superfluid
density, we obtain $I_s/\Icrust \approx 0.94$. For the pasta phases
with anisotropy (rods, plates), we assume that the orientation is
random so that one can average the superfluid density over the three
directions.

\begin{figure}
  \includegraphics[width=8cm]{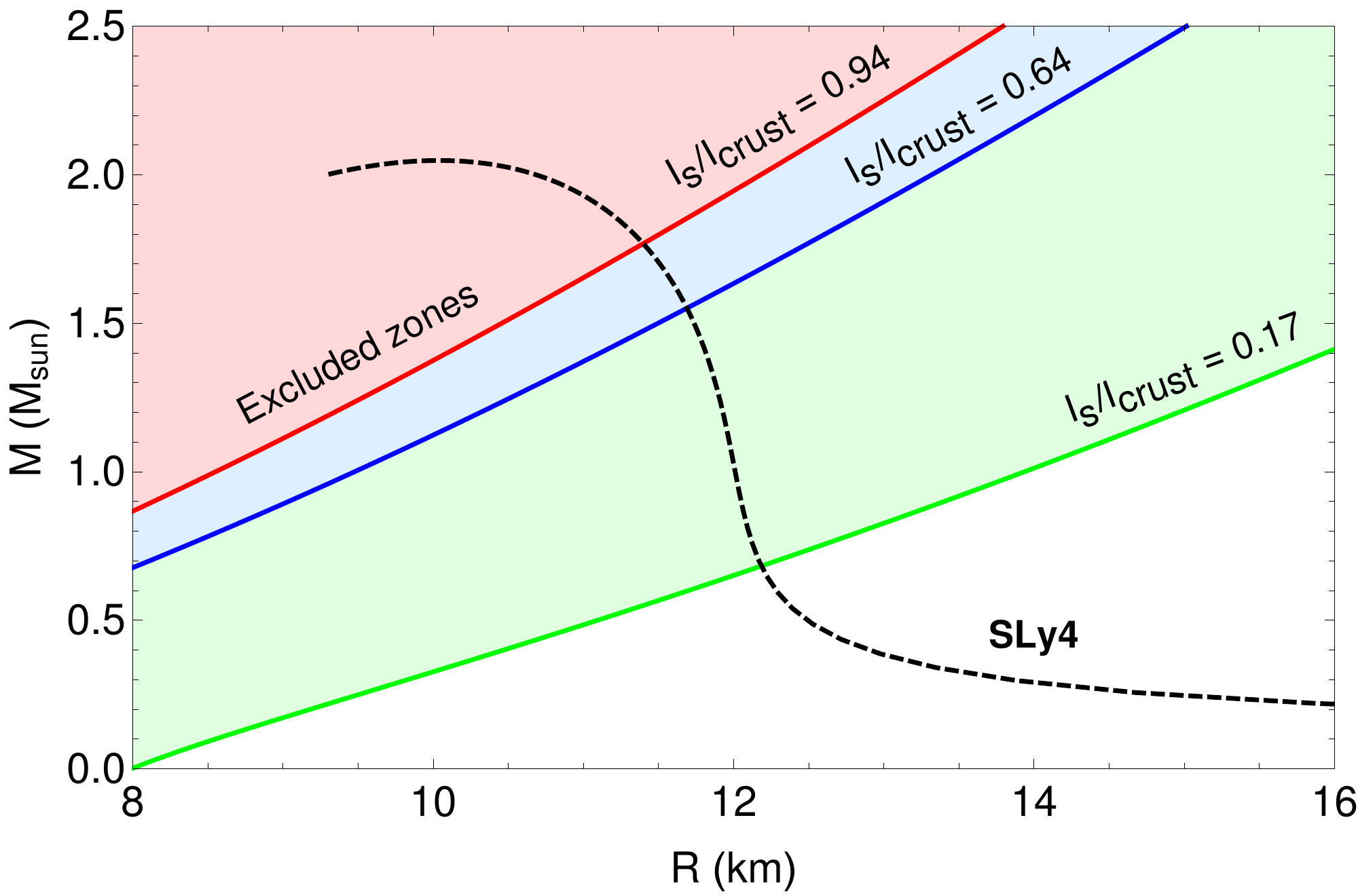}
  \caption{(Color online) Constraints on mass and radius of the Vela
    pulsar from its observed glitch activity for different
      superfluid fractions in the crust: hydrodynamic result
      $I_s/\Icrust = 0.94$ (red), result from band-structure theory
      \cite{Chamel2013PRL} $I_s/\Icrust = 0.17$ (green), and an
      intermediate situation $I_s/\Icrust = 0.64$ (blue) corresponding
      to hydrodynamics in the gas but no superfluidity in the clusters
      (see \Sec{sec:discussion}). We also show as an example the
      mass-radius relation obtained with the SLy4 interaction (dashed
      line). Note that other equations of state would lead to
      different mass-radius relations in a band around the shown one
      (see, e.g., \Ref{SteinerLattimer2013} for an attempt to use
      observational data to constrain the width of this band).
  \label{fig:massradius}}
\end{figure}

For the Vela pulsar, one has $\mathcal{G} \approx 1.6\%$
\cite{Chamel2013PRL}. With the approximations mentioned above
and using our result $I_s/\Icrust \approx 0.94$, this allows
one to identify an excluded zone in the mass-radius diagram,
shown in \Fig{fig:massradius} in red.
Details on the boundary of the excluded zone are given in the
appendix. No assumption has been made so far concerning the EOS in
the core. To give a specific example, we show in
\Fig{fig:massradius} also the mass-radius relation obtained by solving
the Tolman-Oppenheimer-Volkov (TOV) equations
\cite{Tolman1939,Oppenheimer1939} with the EOS given by the SLy4
interaction in the whole star (for the outer crust, we use the results
of \Ref{DouchinHaenselPLB2000}).\footnote{The calculation of
    the outer crust in \Ref{DouchinHaenselPLB2000} and our ETF model
    for the inner crust \cite{MartinUrban2015} are based on the SLy4
    interaction.} One sees that, with this EOS, Vela could have a
mass of up to $\lesssim 1.7\, M_{\text{sun}}$.

The strong entrainment predicted by band-structure theory
results in a much smaller ratio $I_s/\Icrust = 0.17$
\cite{Chamel2013PRL}. As one can see from the corresponding excluded
zone in the mass-radius diagram (zone above the green line in
\Fig{fig:massradius}, this small superfluid fraction can only be
conciliated with the observed glitch activity if either Vela is a very
unusual neutron star with $M < 0.7\, M_{\text{sun}}$
\cite{Chamel2013PRL}, or also the core has a superfluid component that
can serve as a reservoir of angular momentum
\cite{AnderssonGlampedakis2012,Chamel2013PRL}. Another solution to
this puzzle was suggested in \Ref{PiekarewiczFattoyev2014}, where it
was pointed out that uncertainties in the EOS do not exclude the
possibility that the crust could be much thicker than usually assumed.

However, in \Fig{fig:massradius} we see that, with the much larger
superfluid density predicted by our approach, the observed glitch
activity is compatible with the assumption that Vela is a perfectly
normal neutron star, without any need for a superfluid core or an
unusually thick crust. 

\section{Discussion}
\label{sec:discussion}
In this paper, we used a superfluid hydrodynamics approach to
determine how the gas neutrons flow on a microscopic scale around and
through the clusters when the crystal lattice of the clusters is
uniformly and slowly moved through the gas. This allowed us to compute
the densities of superfluid and bound (entrained) neutrons, $\nnsuper$
and $\nnbound$, and the effective mass of the clusters. Surprisingly,
it turned out that $\nnsuper$ is larger than the density of free
neutrons, $\nnfree$. As a consequence, the cluster effective mass
number $\Aeff$ is not only smaller than the number of nucleons that
are spatially located inside the cluster, but even smaller than the
number of energetically bound nucleons.

Our results are in line with those obtained in
\Refs{MagierskiBulgac2004,MagierskiBulgac2004NPA,Magierski2004} using
the same hydrodynamic approach but for the case of an isolated cluster
in an infinite neutron gas. However, in other studies, the opposite
effect was found, namely that the effective mass of the clusters is
increased by the presence of the gas.

For instance, in \Ref{Sedrakian1996}, a hydrodynamic approach was
used, too, but with different boundary conditions at the interface
between the cluster and the gas. In that work, the gas was assumed to
flow around the cluster, increasing the total kinetic energy, while in
our approach and that of
\Refs{MagierskiBulgac2004,MagierskiBulgac2004NPA,Magierski2004} the
permeability of the phase boundary allows the neutrons to flow through
the cluster, reducing the neutron velocity inside the cluster and the
total kinetic energy.

Studies of entrainment in the framework of band-structure theory
\cite{CarterChamel2005,Chamel2012} also predict a strong reduction of
$\nnsuper$ as compared to $\nnfree$, and therefore a strong increase
of $\Aeff$. This approach was developed in analogy to band structure
theory for electrons in condensed-matter physics. However, the
situation of neutrons in the inner crust differs in some respects from
the one of electrons in superconducting metals. In superconductors,
the distance between the energy bands, of the order of a few eV, is
much larger than the pairing gap $\Delta$ which is typically of the
order of a few Kelvin ($10^{-4}$ eV). This is why the pairing affects
only electrons of the conduction band. The spatial extension of a
Cooper pair of electrons is much larger than the unit cell of the
crystal. In contrast, the neutron energy bands in the neutron-star
crust lie very close to one another (cf. Figs. 2--4 in
\Ref{Chamel2012}): for a given quasimomentum $\kv$, there can be many
bands $\alpha$ whose energies $\epsilon_{\alpha\kv}$ are separated by
less than 1 MeV, which is the typical scale for the pairing gap
$\Delta$. This goes along with a coherence length $\xi$ that is
smaller than the unit cell.

For hydrodynamics to be quantitatively accurate, one would need a
coherence length $\xi$ that is also  much smaller than
the clusters. Since this condition is not satisfied, the true answer
lies probably somewhere between the two models, i.e., the
entrainment is maybe stronger than the one predicted by hydrodynamics,
but weaker than the one predicted by band structure theory.

Coming back to the analogy with rotating nuclei which exhibit a
mixture of rotational and irrotational flow as mentioned in
\Sec{sec:model}, one might think about describing the neutrons in the
clusters as a mixture of superfluid neutrons, whose motion is governed
by the phase $\varphi$ of the gap, and normal-fluid neutrons, which
move together with the protons. Recently it was suggested in the
supplemental material of \Ref{WlazlowskiSekizawa2016} to modify the
hydrodynamic model of
\Refs{MagierskiBulgac2004,MagierskiBulgac2004NPA,Magierski2004} in
this sense by reinterpreting the densities $n_{n,\Out}$ and
$n_{n,\In}$ as effective superfluid densities. For instance, if we
assume that all neutrons in the gas but only a fraction $\delta$ of
the neutrons in the cluster participate in the superfluid motion,
\Eq{NeffMagierski} for the effective mass of a single spherical
cluster becomes
\begin{equation}
\Neff = \Nrcluster
  \Big(1-\delta+\frac{(\delta-\gamma)^2}{\delta+2\gamma}\Big) \,.
\label{Neffdelta}
\end{equation}
Analogously, it is straight-forward to generalize also
\Eqs{Neff-spaghetti} and (\ref{nbound-lasagne}) to the case $\delta
< 1$. In the extreme case $\delta=0$ (no superfluidity inside the
clusters, i.e., all neutrons in the cluster move together with the
protons), one retrieves the picture of the gas flowing around the
cluster as in \Ref{Sedrakian1996}, resulting in $\Neff = \Nrcluster
(1+\gamma/2)$. However, this extreme case does not seem to be
realistic, since, e.g., in rotating nuclei at least one half of the
nucleons follow the superfluid motion as one can conclude from the
moments of inertia. Furthermore, we note that the present situation of
a uniform flow of neutrons through the cluster is more favorable for
hydrodynamics than the rotation of nuclei: while in a deformed nucleus
rotating around the $z$ axis the phase $\varphi$ is proportional to
$xy$ \cite{Migdal1959}, our phase is (inside the clusters) only linear
in the coordinates. Therefore, $\delta$ should probably be larger than
one half.

\begin{figure}
  \includegraphics[width=8cm]{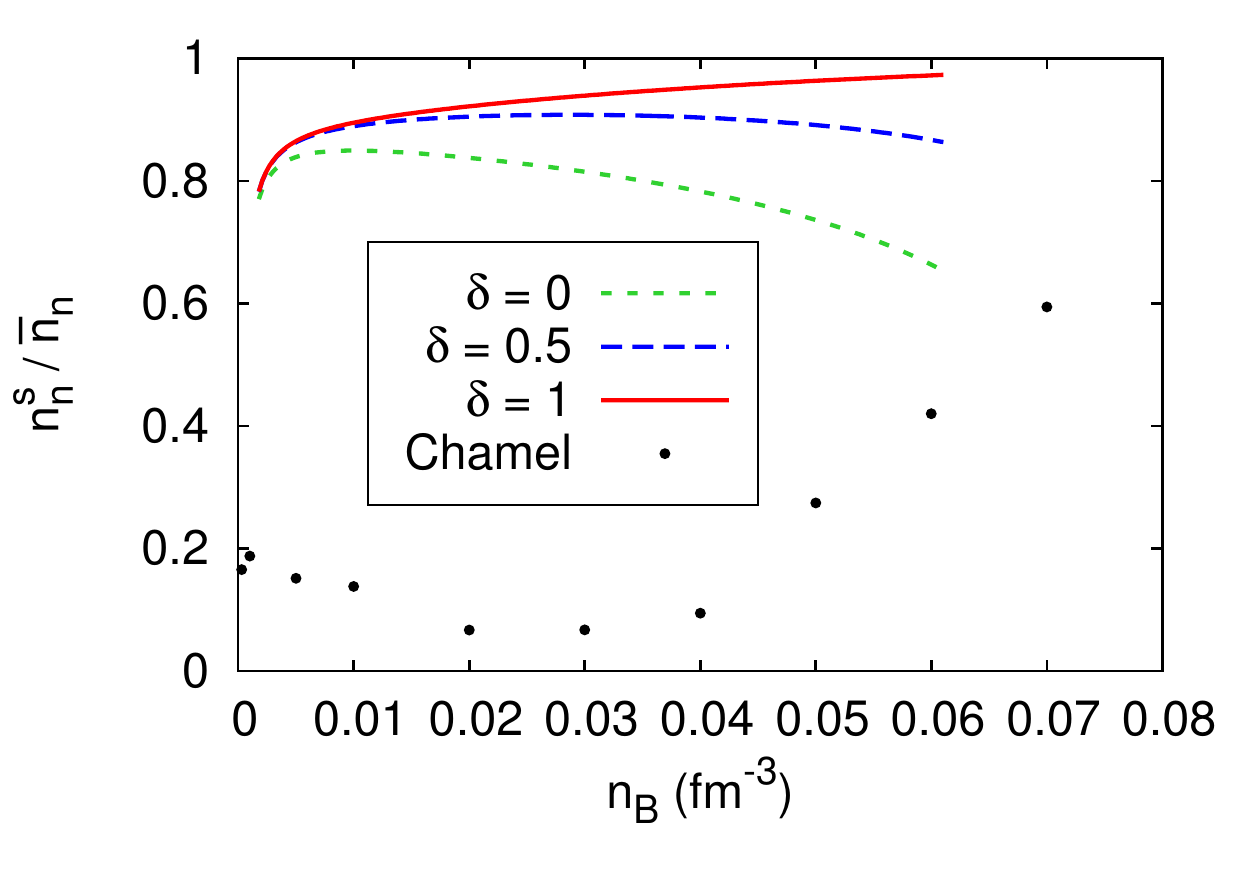}
  \caption{(Color online) Superfluid fraction $\nnsuper/\nnavg$
    as a function of the baryon density $n_B$, obtained under the
    assumption that a fraction $\delta = 0$ (green short dashes),
    $0.5$ (blue long dashes), or $1$ (red solid line) of the neutrons
    in the clusters are superfluid. For comparison, the black circles
    are the result of the band-structure calculations by Chamel
    \cite{Chamel2012}.
  \label{fig:nsnn-ratio-delta}}
\end{figure}

In analogy to the result of \Sec{sec:effmass} that $\Neff$ in the
periodic lattice follows closely the analytic formula
(\ref{NeffMagierski}), we can also compute the superfluid density
$\nnsuper = \nnavg-2\Neff/V_\cell$ with $\Neff$ from
\Eq{Neffdelta}. The resulting superfluid fractions for three values of
$\delta$ are shown in \Fig{fig:nsnn-ratio-delta}.
The case $\delta=1$ corresponds to the one shown already in
\Fig{fig:nsnn-ratio}, but also for $\delta=0.5$ and even in the
extreme case $\delta=0$ we obtain a superfluid density that is
considerably larger than the one of \Ref{Chamel2012}.

Using the superfluid fraction obtained for $\delta = 0$ in
\Eq{eq:IsIcrust}, we find that the superfluid contribution to the
moment of inertia of the crust would still be $I_s/\Icrust \approx
0.64$. The corresponding excluded region in the mass-radius diagram
is the region above the blue line in \Fig{fig:massradius} and,
although it extends to lower masses than the result for $\delta =1$
($I_s/\Icrust \approx 0.94$), it is still compatible with a mass of up
to $\sim 1.5 M_{\text{sun}}$.

In any case, superfluid hydrodynamics remains a strongly simplified model,
not only because of the assumption that the neutron motion is completely
determined by the phase $\varphi$, but also because of the sharp surface of the
clusters. To obtain more reliable results,  one should ideally perform a
QRPA calculation on top of a Hartree-Fock-Bogoliubov (HFB) ground state imposing
the Bloch boundary conditions \cite{AshcroftMermin} on the single-particle wave
functions as in band structure theory. However, at present this objective seems
to be out of reach. Using a much simpler QRPA calculation in a spherical
Wigner-Seitz (WS) cell, as in \Ref{KhanSandulescu2005}, could help to resolve at
least the issues of the effective superfluid density in the cluster and the
most realistic boundary conditions to be used in hydrodynamic calculations.
Instead of the QRPA, one might also use the time-dependent superfluid
local-density approximation (TDSLDA) \cite{Bulgac2013,WlazlowskiSekizawa2016}.
Furthermore, as pointed out in \Ref{KobyakovPethick2013}, one should
probably also consider zero-point oscillations of the clusters that would reduce
the band-structure effects.

\begin{acknowledgments}
We thank Micaela Oertel and Nicolas Chamel for useful
discussions. This work has been funded by the P2IO LabEx
(ANR-10-LABX-0038) in the framework ``Investissements d'Avenir''
(ANR-11-IDEX-0003-01) managed by the French National Research Agency
(ANR).
\end{acknowledgments}
\appendix
\section{Boundary of the excluded zone in the mass-radius diagram}
\label{apx:boundary}
According to \Eq{eq:isi_condition}, the boundary between the allowed
and the excluded zone in \Fig{fig:massradius} corresponds to
$I_s/I=\mathcal{G}$. As mentioned in \Sec{sec:glitches}, we follow
\Ref{Chamel2013PRL} and decompose the ratio $I_s/I$ as
$(I_s/\Icrust)(\Icrust/I)$. The factor $(I_s/\Icrust)$ is given by
\Eq{eq:IsIcrust}. For $(\Icrust/I)$, an analytic expression is given
in Eq.~(47) of \Ref{LattimerPrakash2000}, which can be written in a
compact way as
\begin{equation}
  \frac{\Icrust}{I} = a_0(R)
  \frac{1 - 1.67 \beta - 0.6 \beta^2}
       {2a_1 + 10 a_1 \beta + (1 - 28 a_1) \beta^2} \,.
  \label{eq:prakash}
\end{equation}
In this equation,
\begin{equation}
\beta = \frac{GM}{Rc^2}
\end{equation}
denotes the compactness of the star, with $G$ the gravitational
constant, $M$ the mass of the star, and $R$ its radius. The
dimensionless coefficients $a_i$ that appear in \Eq{eq:prakash} are
given by
\begin{equation}
  a_0(R) = \frac{28 \pi G \Pcore R^2}{3c^4} \,,\quad a_1 =
  \frac{\Pcore}{\ncore m c^2} \,,
\end{equation}
with $\Pcore$ and $\ncore$ the pressure and the density at the
crust-core transition, respectively, and $m$ the neutron mass. Note
that the expression (\ref{eq:prakash}) for $\Icrust/I$ contains $R$
and $M$ as independent variables because no assumption about the EOS
in the core of the star is made, while it depends on the EOS in the
crust through $\Pcore$ and $\ncore$. We use the values corresponding
to our ETF model for the inner crust \cite{MartinUrban2015} based on
the SLy4 interaction: $\Pcore = 0.38$ MeV fm$^{-3}$ and $\ncore =
0.081$ fm$^{-3}$.

For a given radius $R$, the compactness $\beta$ and hence the mass $M$
corresponding to the boundary of the excluded zone shown in
\Fig{fig:massradius} is now obtained as the solution of the quadratic
equation
\begin{multline}
  [0.6+(1 - 28 a_1)b(R)] \beta^2 + [1.67 + 10 a_1 b(R)] \beta\\
    + [1 + 2 a_1 b(R)] = 0 \,,
\end{multline}
where $b(R)$ is defined as
\begin{equation}
  b(R) = \frac{\mathcal{G}}{a_0(R)} \left( \frac{I_s}{\Icrust} \right)^{-1}\,.
\end{equation}

\nocite*
\bibliographystyle{apsrev4-1}
\bibliography{hydro}

\end{document}